\documentclass[12pt,onecolumn]{IEEEtran}
\usepackage{mathrsfs,pslatex}
\usepackage{amsfonts}
\usepackage{indentfirst}
\usepackage{amsmath}
\usepackage{amssymb}
\usepackage{color}

\newtheorem{Theorem}{Theorem}
\newtheorem{Corollary}{Corollary}

\newtheorem{Remark}{Remark}
\newtheorem{Lemma}{Lemma}
\newtheorem{Example}{Example}
\newtheorem{Open}{Open Problem}

\newcommand{\fpm}{{\mathbb F}_{p^m}}
\newcommand{\fthreem}{{\mathbb F}_{3^m}}
\newcommand{\fp}{{\mathbb F}_{p}}
\newcommand{\fthree}{{\mathbb F}_{3}}
\newcommand{\fq}{{\mathbb F}_{q}}

\begin{document}
%
\title{Optimal Ternary Cyclic Codes with Minimum Distance Four and Five}

\author{Nian Li,
       Chunlei Li,
       Tor Helleseth, 
       Cunsheng Ding, 
     and  Xiaohu Tang  
 \thanks{N. Li and X. Tang are with the Information Security and National Computing Grid Laboratory, Southwest Jiaotong
         University, Chengdu, 610031, China (e-mail: nianli.2010@gmail.com, xhutang@swjtu.edu.cn). N. Li is current a visiting Ph. D student (Sept.
         2011- Aug. 2013) in the Department of Informatics, University of Bergen.}
 \thanks{C. Li and T. Helleseth are with the Department of Informatics, University of Bergen, N-5020 Bergen, Norway
         (e-mail: chunlei.li@ii.uib.no, tor.helleseth@ii.uib.no).}
  \thanks{C. Ding is with the Department of Computer Science and Engineering, The Hong Kong University of Science and Technology, Clear Water Bay, Kowloon, Hong Kong, China (e-mail: cding@ust.hk).}
}

\date{}

\maketitle

\begin{abstract}
Cyclic codes are an important subclass of linear codes and have wide applications in data storage systems, communication systems and consumer electronics. In this paper, two families of optimal ternary cyclic codes are presented.
The first family of cyclic codes has parameters $[3^m-1, 3^m-1-2m, 4]$ and contains a class of conjectured cyclic codes
and several new classes of optimal cyclic codes.
The second family of cyclic codes has parameters $[3^m-1, 3^m-2-2m, 5]$ and contains a number of classes of cyclic codes
that are obtained from perfect nonlinear functions over $\fthreem$, where $m>1$ and is a positive integer.
\end{abstract}


\IEEEpeerreviewmaketitle

\section{Introduction}

Throughout this paper let $p$ and $m$ be a prime and a positive integer respectively, and let $\fpm$ denote the finite field with $p^m$ elements.  An $[n,k,d]$ linear code $\mathcal{C}$ over $\fp$ is a $k$-dimensional subspace of $\fp^n$ with minimum Hamming
distance $d$, and is called {\em cyclic} if any cyclic shift of a codeword
is another codeword of $\mathcal{C}$. By identifying  $(c_0,c_1,\cdots,c_{n-1})\in \mathcal{C}$ with
 \[c_0+c_1x+c_2x^2+\cdots+c_{n-1}x^{n-1}\in\fp[x]/(x^n-1),\]
 any cyclic code of length $n$ over $\fp$ corresponds to an ideal of the polynomial residue class ring $\fp[x]/(x^n-1)$.
 Note that every ideal of $\fp[x]/(x^n-1)$ is principal. Any cyclic code $\mathcal{C}$ can be expressed as $\mathcal{C}=\langle g(x) \rangle$, where $g(x)$ is monic and has the least degree. This polynomial  $g(x)$ is called the {\em generator polynomial} and $h(x)=(x^n-1)/g(x)$ is referred to as the {\em parity-check polynomial} of $\mathcal{C}$.

 Cyclic codes are an important subclass of linear codes and have been extensively studied (see for example \cite{Betti-Sala},  \cite{Ding-Ling}, \cite{Ding-Yang}, \cite{Feng07}, \cite{TFeng}, \cite{Jia-Ling-Xing}, \cite{LuoFeng2}, \cite{Zeng-HJYC} and \cite{Zeng-SH} for some recent developments).
 Let $\alpha$ be a generator of $\fthreem^*=\fthreem\backslash\{0\}$ and let $m_{\alpha^i}(x)$ denote the minimal polynomial of $\alpha^i$ over $\fthree$. A class of cyclic codes $\mathcal{C}_{(1,e)}$ over $\fthree$ with generator polynomial $m_{\alpha}(x)m_{\alpha^e}(x)$, where $1\leq e\leq 3^m-1$ and $e$ is not in the 3-cyclotomic coset modulo $3^m-1$ containing $1$, was investigated in \cite{Carlet-Ding-Yuan} and \cite{Ding-Helleseth}. Carlet, Ding and Yuan proved that the code $\mathcal{C}_{(1,e)}$ has parameters $[3^m-1,3^m-1-2m,4]$ when $x^e$ are certain perfect nonlinear (PN) monomials \cite{Carlet-Ding-Yuan}. Employing some monomials $x^e$ over $\fthreem$, including almost perfect nonlinear (APN) monomials, Ding and Helleseth \cite{Ding-Helleseth} obtained several classes of ternary cyclic codes with  parameters $[3^m-1,3^m-1-2m,4]$ which are optimal according to the Sphere Packing bound.  In addition, nine open problems about this kind of optimal ternary cyclic codes were proposed in \cite{Ding-Helleseth}.  Notably, as a class of subcodes of $\mathcal{C}_{(1,e)}$, the cyclic codes with generator polynomial $(x-1)m_{\alpha}(x)m_{\alpha^e}(x)$, which are denoted by  $\mathcal{C}_{(0,1,e)}$, were investigated in \cite{Carlet-Ding-Yuan} and \cite{YCD06}, and it was proven in \cite{Carlet-Ding-Yuan} that $\mathcal{C}_{(0,1,e)}$ has parameters  $[3^m-1,3^m-2m-2,5]$ if $x^e$ is PN.

In this paper, we will present a number of classes of new optimal ternary cyclic codes with parameters $[3^m-1,3^m-1-2m,4]$
and $[3^m-1,3^m-2-2m,5]$.  We will first settle an open problem proposed in \cite{Ding-Helleseth} and then construct several classes of new optimal ternary cyclic codes with parameters $[3^m-1,3^m-1-2m,4]$ using some monomials over $\fthreem$.  We then derive a number of classes of optimal ternary cyclic codes with parameters $[3^m-1,3^m-2-2m,5]$ by considering the subcodes of $\mathcal{C}_{(1,e)}$ with generator polynomial $(x+1)m_{\alpha}(x)m_{\alpha^e}(x)$ over $\fthree$. Following the notations in \cite{Ding-Helleseth}, we denote  by $\mathcal{C}_{(1,e,s)}$ the cyclic code with generator polynomial $(x+1)m_{\alpha}(x)m_{\alpha^e}(x)$, where $s=\frac{3^m-1}{2}$. It will be shown in this paper that the ternary cyclic code $\mathcal{C}_{(1,e,s)}$ has parameters $[3^m-1,3^m-2m-2,5]$ and is optimal for several classes of properly chosen integers $e$. The optimality of $\mathcal{C}_{(1,e,s)}$ is established by virtue of properties of PN functions over $\fthreem$.

\section{Auxiliary results about cyclotomic cosets, the codes $\mathcal{C}_{(1,e)}$ and polynomials}\label{sec-basic}

A function $f$ from $\fpm$ to itself is called  {\em perfect nonlinear} (PN) or {\em planar} if
  \[\max_{0\ne a\in\fpm}\max_{b\in\fpm}|\{x\in\fpm: f(x+a)-f(x)=b\}|=1,\]
and {\em almost perfect nonlinear} (APN) if
  \[\max_{0\ne a\in\fpm}\max_{b\in\fpm}|\{x\in\fpm: f(x+a)-f(x)=b\}|=2.\]
In this paper we will need the notions of PN and APN functions \cite{Helleseth-Rong-S,Zha}.

For a prime $p$, the $p$-cyclotomic coset modulo $p^m-1$ containing $j$ is defined as
\[C_j=\{jp^s \bmod{(p^m-1)}: s=0,1,\cdots,m-1\}.\]

The following lemma will be frequently used in the sequel.

\begin{Lemma}\label{lem-zero} {\rm (\cite{Ding-Helleseth}) }
For any $1 \le e \le p^m-2$ with $\gcd(e, p^m-1)=2$, the length of the $p$-cyclotomic
coset $C_e$ is equal to $m$.
\end{Lemma}

 Ding and Helleseth proved the following fundamental theorem about the ternary codes $\mathcal{C}_{(1,e)}$.

\begin{Theorem}{\rm(\cite[Thm. 4.1]{Ding-Helleseth})}\label{thm-Ding-Helleseth}
Let $e\not\in C_1$ and $|C_e|=m$. The ternary cyclic code $\mathcal{C}_{(1,e)}$ has parameters $[3^m-1,3^m-1-2m,4]$ if and only if the following conditions are satisfied:
\begin{itemize}
     \item[C1:] $e$ is even;
     \item[C2:] the equation $(x+1)^e+x^e+1=0$ has the only solution $x=1$ in $\fthreem$; and
     \item[C3:] the equation $(x+1)^e-x^e-1=0$ has the only solution $x=0$ in $\fthreem$.
\end{itemize}
\end{Theorem}

We shall need the following lemma in the sequel, in addition to Theorem \ref{thm-Ding-Helleseth}.

\begin{Lemma}\label{lem-roots-irre-poly} {\rm (\cite[Thm. 2.14]{Lidl-N})}
Let $q$ be a prime power and let $f(x)$ be an irreducible polynomial over $\fq$ of degree $n$. Then $f(x)=0$ has a root $x$ in $\mathbb{F}_{q^n}$. Furthermore, all the roots of $f(x)=0$ are simple and are given by the $n$ distinct elements $x$, $x^q$, $x^{q^2}$, $\cdots$, $x^{q^{n-1}}$ of $\mathbb{F}_{q^n}$.
 \end{Lemma}

 Let us take $f(x)=x^3+x^2+x-1\in\fthree[x]$ as an example to show how Lemma \ref{lem-roots-irre-poly} works.
 Note that $f(0)=f(1)=2\ne 0$ and $f(2)=f(-1)=1\ne 0$. This means that $f(x)=x^3+x^2+x-1$ is a cubic irreducible
 polynomial over $\fthree[x]$. Then by Lemma \ref{lem-roots-irre-poly}, $f(x)=0$ has no solutions in $\fthreem$
 if and only if $m\not\equiv 0 \pmod{3}$. This idea will be frequently employed in the sequel to prove some of the
 main results of this paper.

For any given $f(x)\in\fthree[x]$, if one factorizes $f(x)$ over $\fthree$, then the number of solutions of $f(x)=0$ in $\fthreem$ can be determined with Lemma \ref{lem-roots-irre-poly}. However, the factorization of a polynomial is
normally a hard problem. In this paper, we mainly consider the cyclic code $\mathcal{C}_{(1,e)}$ for special values of $e$,
where only low-degree polynomials over $\fthree[x]$ should be factorized. In fact, to apply Lemma \ref{lem-roots-irre-poly}, sometimes one only needs to know the degrees of the irreducible factors of $f(x)$.

 The following lemmas are basic results about polynomials over finite fields and will be employed in the sequel.

\begin{Lemma}\label{lem-Division-poly} {\rm (\cite{Lidl-N})}
Let $q$ be a prime power and $g(x)$ be a polynomial in $\fq[x]$. Then for any $f(x)\in\fq[x]$ there exist polynomials $h(x), r(x)\in\fq[x]$ such that $f(x)=g(x)h(x)+r(x)$, where $\deg(r(x))<\deg(g(x))$. Moreover, $\gcd(f(x), g(x))=\gcd(g(x),r(x))$.
\end{Lemma}

\begin{Lemma}\label{lem-all-ir-poly}  {\rm (\cite[Thm. 3.20]{Lidl-N})}
For every finite field $\fq$ and every positive integer $n$, where $q$ is a prime power, the product of all monic irreducible polynomials over $\fq[x]$ whose degrees divide $n$ is equal to $x^{q^n}-x$.
\end{Lemma}

For a given $f(x)\in\fq[x]$ with low degree, Lemmas \ref{lem-Division-poly} and \ref{lem-all-ir-poly} can be used
to determine the degrees of irreducible factors of $f(x)$. For example, let $f(x)=x^8+x^7-x^6+x^4-x^3+x^2-1\in\fthree[x]$.
Applying Lemma \ref{lem-Division-poly}, one gets that $\gcd(f(x),x^{3^3}-x)=1$ and $\gcd(f(x),x^{3^4}-x)=x^2+x-1$. It then follows from Lemma \ref{lem-all-ir-poly} that $f(x)$ has the irreducible factor $x^2+x-1$ but no irreducible factor with degree equals to $1$, $3$ and $4$. This implies that $f(x)$ has an irreducible factor with degree $6$.

\section{Solving an open problem about the ternary cyclic codes $\mathcal{C}_{(1,e)}$}\label{sec-open}

With the preparations in Section \ref{sec-basic}, in this section we settle the following open problem proposed in \cite{Ding-Helleseth}:

\begin{Open}{\rm(\cite{Ding-Helleseth})}\label{open-one}
Let $e=2(3^{m-1}-1)$. Does the ternary cyclic code $\mathcal{C}_{(1,e)}$ have parameters $[3^m-1,3^m-1-2m,4]$ if $m\geq 5$ and $m$ is prime?
\end{Open}

To solve this problem, we need to prove the following lemmas.

\begin{Lemma}\label{lem-open-Ce}
Let $m$ be odd and $e=2(3^{m-1}-1)$. Then $|C_e|=m$ and $C_1 \cap C_e = \emptyset$.
\end{Lemma}

\begin{IEEEproof}
It is easily seen that $\gcd(e, 3^m-1)=2$. It then follows from Lemma \ref{lem-zero} that $|C_e|=m$.
Since both $e$ and $3^m-1$ are even, it is obvious that $C_1 \cap C_e = \emptyset$.
This completes the proof.
\end{IEEEproof}

After proving Lemma \ref{lem-open-Ce}, we now consider Conditions C2 and C3 in Theorem \ref{thm-Ding-Helleseth}
for $e=2(3^{m-1}-1)$.

\begin{Lemma}\label{lem-open-C2}
Let $e=2(3^{m-1}-1)$. Then Condition C2 in Theorem \ref{thm-Ding-Helleseth} is met if and only if $m\not\equiv 0 \pmod{3}$.
\end{Lemma}

\begin{IEEEproof}
Note that $e$ is even and $x=0$ is not a solution of $(-x-1)^e+x^e+1=0$. Then Condition C2 is satisfied if and only if $(x+1)^e+x^e+1=0$ has the only solution $x=1$ in $\fthreem$. Raising both sides of this equation to the power of 3
gives $(x+1)^{3e}+x^{3e}+1=0$. Note that $x \ne 0$, $x+1\ne 0$ and $3e=2(3^m-3)=2(3^m-1)-4$. Then the equation
$(x+1)^e+x^e+1=0$ is equivalent to $(x+1)^{-4}+x^{-4}+1=0$, i.e.,
\begin{equation}\label{Eq-open-C2}
    (x+1)^4x^4+(x+1)^4+x^4=0.
\end{equation}
Denote $f(x)=(x+1)^4x^4+(x+1)^4+x^4=x^8+x^7+x^5+x^3+x+1$. Applying Lemma \ref{lem-Division-poly}, one gets that
$\gcd(f(x),x^3-x)=x-1$, $\gcd(f(x),x^{3^2}-x)=x-1$, and $\gcd(f(x),x^{3^3}-x)=x^7-x^6-x^5+x^2+x-1$. It then follows from Lemma \ref{lem-all-ir-poly} and $\deg(f(x))=8$ that $f(x)$ has the two cubic irreducible factors $x^3 + x^2 + x + 2$ and $x^3 + 2x^2 + 2x + 2$ over $\fthree$ and the factor $(x-1)^2$. Thus (\ref{Eq-open-C2})  has the only solution $x=1$ in $\fthreem$ if and only if  $m\not\equiv 0 \pmod{3}$ by Lemma \ref{lem-roots-irre-poly}.
This completes the proof.
\end{IEEEproof}

\begin{Lemma}\label{lem-open-C3}
Let $e=2(3^{m-1}-1)$. Then Condition C3 in Theorem \ref{thm-Ding-Helleseth}  is met if and only if $m$ is odd and $m\not\equiv 0 \pmod{3}$.
\end{Lemma}

\begin{IEEEproof}
Let $x \in \fthreem$ be a solution of $(x+1)^e-x^e-1=0$. Then we have $(x+1)^{3e}-x^{3e}-1=0$. Notice that $3e=2(3^m-3)=2(3^m-1)-4$.
Then $(x+1)^{3e}-x^{3e}-1=0$ has the only solution $x=0$ if and only if $(x+1)^{-4}-x^{-4}-1=0$ has no
solution in $\fthreem^*$. Multiplying both sides of this equation with $x^4(x+1)^4$ gives
\begin{equation}\label{Eq-open-C3}
    (x+1)^4x^4+(x+1)^4-x^4=0.
\end{equation}
Therefore Condition C3 is equivalent to showing that (\ref{Eq-open-C3}) has no solutions in $\fthreem$. Denote $g(x)=(x+1)^4x^4+(x+1)^4-x^4=x^8+x^7+x^5+x^4+x^3+ x+1$. By Lemma \ref{lem-Division-poly}, we have $\gcd(g(x),x^3-x)=1$, $\gcd(g(x),x^{3^2}-x)=x^2+1$, and $\gcd(g(x),x^{3^3}-x)=x^6+x^5-x^4-x^2+x+1$.
It then follows from Lemma \ref{lem-all-ir-poly} that $g(x)$ has the irreducible factor $x^2+1$ and the two cubic
irreducible factors $x^3 + 2x + 2$ and $x^3 + x^2 + 2$. Then the desired conclusion follows from Lemma
\ref{lem-roots-irre-poly}. This completes the proof.
\end{IEEEproof}

The answer to Open Problem \ref{open-one} is given in the following theorem.

\begin{Theorem}\label{thm-Answer}
Let $m$ be odd, $m\not\equiv 0 \pmod{3}$ and $e=2(3^{m-1}-1)$. Then the ternary cyclic code $\mathcal{C}_{(1,e)}$ has parameters $[3^m-1,3^m-1-2m,4]$.
\end{Theorem}

\begin{IEEEproof}
The conclusions follow from  Lemmas \ref{lem-open-Ce}-\ref{lem-open-C3} and Theorem \ref{thm-Ding-Helleseth}.
\end{IEEEproof}

\begin{Example}
Let $m=5$ and let $\alpha$ be the generator of $\fthreem^*$ with $\alpha^5+2\alpha+1=0$.  Then the code $\mathcal{C}_{(1,e)}$ of Theorem \ref{thm-Answer} has parameters $[242, 232,4]$ and generator polynomial $x^{10}+2x^9+x^8+x^5+x^4+x^3+2x^2+2x+2$.
\end{Example}

\section{New optimal ternary cyclic codes with parameters  $[3^m-1,3^m-1-2m,4]$}

Inspired by the idea for solving Open Problem \ref{open-one}, we construct new optimal ternary cyclic codes with
parameters  $[3^m-1,3^m-1-2m,4]$ using other monomials $x^e$ over $\fthreem$ in this section.

 \subsection{The first class of optimal ternary cyclic codes with parameters  $[3^m-1,3^m-1-2m,4]$}

In this subsection, we consider the exponents $e$ of the form $e=\frac{3^m-1}{2}-r$, where $r$ and $m$ have the same parity. Denote the quadratic character of $\fthreem$ by $\eta$ which is defined by $\eta(0)=0$,  $\eta(x)=1$ if $x$ is a nonzero square in  $\fthreem$ and $\eta(x)=-1$ if $x$ is a nonzero nonsquare in $\fthreem$.  Note that $e$ is even. Then
Condition C2 in Theorem \ref{thm-Ding-Helleseth} is satisfied if and only if $(x+1)^e+x^e+1=0$ has the only solution $x=1$ in $\fthreem$. With the quadratic character of $\fthreem$, $(x+1)^e+x^e+1=0$ can be written as $\eta(x+1)(x+1)^{-r}+\eta(x)x^{-r}+1=0$. Multiplying with $x^r(x+1)^r$ both sides of this equation yields that
\begin{equation}\label{Eq-A-C2}
(x+1)^rx^r+\eta(x+1)x^r+\eta(x)(x+1)^r=0.
\end{equation}
Then C2 is satisfied if and only if  (\ref{Eq-A-C2}) has the only solution $x=1$ with $\eta(x(x+1))\ne 0$ since neither $x=0$ nor $x=-1$ are the solutions of $(x+1)^e+x^e+1=0$. Similarly, one can conclude that C3 is satisfied if and only if
  \begin{equation}\label{Eq-A-C3}
    (x+1)^rx^r+\eta(x)(x+1)^r-\eta(x+1)x^r=0
  \end{equation}
has no solution $x$ in $\fthreem$ with $\eta(x(x+1))\ne 0$.

The following theorem then follows from Theorem \ref{thm-Ding-Helleseth} and the foregoing discussions.

\begin{Theorem}\label{thm-class-A}
Let $e=\frac{3^m-1}{2}-r$, $e\not\in C_1$ and $|C_e|=m$, where $r$ and $m$ have the same parity. Then the ternary cyclic code $\mathcal{C}_{(1,e)}$ has parameters $[3^m-1,3^m-1-2m,4]$ if (\ref{Eq-A-C2}) has the only solution $x=1$ and (\ref{Eq-A-C3}) has no nonzero solution $x$ in $\fthreem$ with $\eta(x(x+1))\ne 0$ .
\end{Theorem}

As in Lemmas \ref{lem-open-C2} and \ref{lem-open-C3}, the solutions of (\ref{Eq-A-C2}) and (\ref{Eq-A-C3}) can be similarly discussed for a given $r$.

\begin{Corollary}\label{cor-A-1}
Let $m\equiv 2 \pmod{4}$ and $e=\frac{3^m-1}{2}-2$. Then the ternary cyclic code $\mathcal{C}_{(1,e)}$ has parameters $[3^m-1,3^m-1-2m,4]$.
\end{Corollary}

\begin{IEEEproof}
Notice that $3^m-1\equiv 0 \pmod{8}$ as $m$ is even. It then follows that $e\equiv 2 \pmod{4}$ and  $\gcd(e,3^m-1)=\gcd(e,4)=2$. By Lemma \ref{lem-zero} we have $|C_e|=m$.
On the other hand, $e\not\in C_1$ since $e$ is even. For $r=2$,  we will discuss (\ref{Eq-A-C2})
by distinguishing among the following cases:
\begin{enumerate}
\item  $(\eta(x),\eta(x+1))=(1,1)$: In this case, (\ref{Eq-A-C2}) is reduced to $x^4-x^3-x+1=(x-1)^4=0$, i.e.,
it has the only solution $x=1$ since $\eta(1)=1$ and $\eta(1+1)=\eta(-1)=1$ for even $m$.

\item  $(\eta(x),\eta(x+1))=(1,-1)$: In this case, (\ref{Eq-A-C2}) is simplified to $x^4-x^3+x^2-x+1=0$.
Applying Lemma \ref{lem-Division-poly}, one obtains that $\gcd(x^4-x^3+x^2-x+1,x^{3^2}-x)=1$,
which implies that $x^4-x^3+x^2-x+1$ has no irreducible factors of degrees 1 and 2 by Lemma \ref{lem-all-ir-poly}.
Therefore $x^4-x^3+x^2-x+1$ is irreducible over $\fthree$.

\item  $(\eta(x),\eta(x+1))=(-1,1)$: Similar as in Case 2), in this case (\ref{Eq-A-C2}) is reduced to $x^4-x^3+x^2+x-1=0$,
which is irreducible over $\fthree$.

\item $(\eta(x),\eta(x+1))=(-1,-1)$: In this case one can similarly prove that $x^4-x^3-x^2+x-1$ is irreducible over $\fthree$.
\end{enumerate}
Therefore, by Lemma \ref{lem-roots-irre-poly}, (\ref{Eq-A-C2}) has the only solution $x=1$ if
$m\equiv 2 \pmod{4}$. It can be similarly proved that (\ref{Eq-A-C3}) has no solution $x$ in $\fthreem$
with $\eta(x(x+1)) \ne 0$. Then the desired conclusions follow from Theorem \ref{thm-class-A}.
This completes the proof.
\end{IEEEproof}

\begin{Example}
Let $m=6$ and let $\alpha$ be the generator of $\fthreem^*$ with $\alpha^6+2\alpha^4+\alpha^2+2\alpha+2$.
Then the code $\mathcal{C}_{(1,e)}$ of Corollary \ref{cor-A-1} has parameters $[728, 716,4]$ and generator polynomial
$x^{12}+2x^{10}+x^9+x^8+2x^5+x^4+x^3+2x^2+2$.
\end{Example}

 Notice that $e=\frac{3^m-1}{2}-1=\frac{3^m-3}{2}$ and $e=\frac{3^m-1}{2}-3\equiv 3\cdot(\frac{3^m-1}{2}-1)
 \pmod{3^m-1}$, i.e., the two cases that $r=1$ and $r=3$ are covered by Theorem 6.1 in \cite{Ding-Helleseth} since $x^{\frac{3^m-3}{2}}$ is an almost perfect nonlinear function in $\fthreem$ if $m\geq 5$ and $m$ is odd \cite{Helleseth-Rong-S}. In the following, we consider the case that $r=5$.

\begin{Corollary}\label{cor-A-2}
Let $m$ be odd and $e=\frac{3^m-1}{2}-5$. Then the ternary cyclic code $\mathcal{C}_{(1,e)}$ has parameters $[3^m-1,3^m-1-2m,4]$.
\end{Corollary}

\begin{IEEEproof}
Since $\gcd((3^m-1)/2,5)=1$ for odd $m$,  $\gcd(3^m-1,e)=\gcd(e,10)=2$. It then follows from Lemma \ref{lem-zero} that $|C_e|=m$.  Since $e$ is even, $e \not\in C_1$. For $r=5$, we below discuss only the solution $x$ of (\ref{Eq-A-C2}) with  $\eta(x(x+1))\ne 0$ since that of (\ref{Eq-A-C3}) can be dealt with in the same manner.
\begin{enumerate}
\item $(\eta(x),\eta(x+1))=(1,1)$: In this case, (\ref{Eq-A-C2}) is reduced to $f(x)=x^{10}-x^9+x^8+x^7-x^6-x^4+x^3+x^2-x+1=0$. Applying Lemma \ref{lem-Division-poly}, one obtains that $\gcd(f(x),x^{3^i}-x)=1$ for
all $i \in \{1,2,3,4,5\}$. It then follows from Lemma \ref{lem-all-ir-poly} that $f(x)$ is irreducible over $\fthree$.

\item  $(\eta(x),\eta(x+1))=(1,-1)$: In this case (\ref{Eq-A-C2}) becomes $f(x)=x^{10}-x^9+x^8+x^7-x^6+x^5-x^4+x^3+x^2-x+1=0$. Applying Lemma \ref{lem-Division-poly} one gets that
$\gcd(f(x),x^3-x)=x-1$, $\gcd(f(x),x^{3^2}-x)=x^7-x^6+x^5-x^4+x^3-x^2+x-1$ and $\gcd(f(x),x^{3^3}-x)=x-1$.
It then follows from Lemma \ref{lem-all-ir-poly} that $f(x)$ has the factor $(x-1)^4$ and three quadratic irreducible
factors (they are $x^2 + 1$, $x^2 + x + 2$ and $x^2 + 2x + 2$). When $m$ is odd,  $\eta(1)=1$ and $\eta(1+1)=\eta(-1)=-1$, i.e., $x=1$ is indeed a solution of  (\ref{Eq-A-C2}).

\item $(\eta(x),\eta(x+1))=(-1,1)$: In this case (\ref{Eq-A-C2}) is reduced to $f(x)=x^{10}-x^9+x^8+x^7-x^6+x^5+x^4-x^3-x^2+x-1=0$.  Similar as in Case 1), one can prove that $f(x)$ is irreducible over $\fthree$.

\item $(\eta(x),\eta(x+1))=(-1,-1)$: In this case one can similarly prove that $x^{10}-x^9+x^8+x^7-x^6-x^5+x^4-x^3-x^2+x-1$ is irreducible over $\fthree$.
\end{enumerate}

Since $m$ is odd, $m\not\equiv 0 \pmod{10}$. It then follows from Lemma \ref{lem-roots-irre-poly} that
(\ref{Eq-A-C2}) has the only solution $x=1$ in $\fthreem$ with $\eta(x(x+1))\ne 0$. Then the desired conclusions
follow from Theorem \ref{thm-class-A}. This completes the proof.
\end{IEEEproof}

\begin{Example}
Let $m=5$ and let $\alpha$ be the generator of $\fthreem^*$ with $\alpha^5+2\alpha+1=0$.  Then the code $\mathcal{C}_{(1,e)}$ of Corollary \ref{cor-A-2} has parameters $[242, 232,4]$ and generator polynomial $x^{10}+2x^9+x^8+2x^7+x^6+x^5+x^4+2x^3+2$.
\end{Example}

\begin{Remark}
By Lemmas \ref{lem-Division-poly} and \ref{lem-all-ir-poly}, more new optimal ternary codes can also be obtained
from other values of $r$, for example, $r=7, 10, 11$. It should be noted that $e=\frac{3^m-15}{2}$ if $r=7$, which is equivalent to $e=\frac{3^{m-1}-5}{2}$. This is a special case of Open Problem 7.10 in \cite{Ding-Helleseth}.
\end{Remark}

 \subsection{The second class of optimal ternary cyclic codes with parameters  $[3^m-1,3^m-1-2m,4]$}

The ternary cyclic code $\mathcal{C}_{(1,e)}$ for $e=\frac{3^m-1}{2}+r$, where $r$ and $m$ have the same parity, is considered in this subsection.  With similar discussions for (\ref{Eq-A-C2}) and  (\ref{Eq-A-C3}) conducted in the preceding
subsection, one can prove the following theorem with Theorem \ref{thm-Ding-Helleseth}.

\begin{Theorem}\label{thm-class-B}
Let $e=\frac{3^m-1}{2}+r$, $e\not\in C_1$ and $|C_e|=m$, where $r$ and $m$ have the same parity. Then the ternary cyclic code $\mathcal{C}_{(1,e)}$ has parameters $[3^m-1,3^m-1-2m,4]$ if
     \begin{equation}\label{Eq-B-C2}
     \eta(x+1)(x+1)^r+\eta(x)x^r+1=0
     \end{equation}
    has the only solution $x=1$ in $\fthreem$ and
   \begin{equation}\label{Eq-B-C3}
     \eta(x+1)(x+1)^r-\eta(x)x^r-1=0
   \end{equation}
   has no nonzero solution in $\fthreem$.
\end{Theorem}

Using Theorem \ref{thm-class-B}, one can verify that $\mathcal{C}_{(1,e)}$ with $e=\frac{3^m-1}{2}+r$  for
$r \in \{1,2,\cdots,6\}$ either is not optimal or has been treated in \cite{Ding-Helleseth}. Thus we start with
$r=7$ below.

 \begin{Corollary}\label{cor-B-1}
 Let $m$ be odd and $e=\frac{3^m-1}{2}+7$. Then the ternary cyclic code $\mathcal{C}_{(1,e)}$ has parameters $[3^m-1,3^m-1-2m,4]$.
\end{Corollary}

\begin{IEEEproof}
 Since $m$ is odd, $e$ is even. Thus $e\not\in C_1$.  It is easily verified that $e\equiv 2 \pmod{3}$ and $\gcd(\frac{3^m-1}{2}, 7)=1$. Hence $\gcd(e,3^m-1)=\gcd(e,14)=2$. It then follows from Lemma \ref{lem-zero} that $|C_e|=m$. In what follows, we prove that (\ref{Eq-B-C2}) has the only solution $x=1$ in $\fthreem$ for $r=7$. This is done by distinguishing
among the following cases:
\begin{enumerate}
\item $(\eta(x),\eta(x+1))=(1,1)$: In this case, (\ref{Eq-B-C2}) is reduced to $f(x)=2x^7+x^6+2x^4+2x^3+x+2=0$.
It is easily checked that $\gcd(f(x),x^{3^2}-x)=(x+1)(x^2+1)$.  By Lemma \ref{lem-all-ir-poly}, $f(x)$ has the factor
$(x+1)(x^2+1)$ and an irreducible factor of degree $4$ (i.e., $x^4 + x^3 + x^2 + x + 1$). Note that $x=-1$ is not
a solution of (\ref{Eq-B-C2}).

\item  $(\eta(x),\eta(x+1))=(1,-1)$: In this case (\ref{Eq-B-C2}) is simplified to $f(x)=2x^6+x^4+x^3+2x=0$.
It is easily verified that $f(x)=x(x+1)(x-1)^4$. Note that $x=1$ is indeed a solution of (\ref{Eq-B-C2}) since $\eta(1+1)=\eta(-1)=-1$ for odd $m$.

\item  $(\eta(x),\eta(x+1))=(-1,1)$: In this case, (\ref{Eq-B-C2}) is reduced to $f(x)=x^6+2x^4+2x^3+x+2=0$.
Similar as in Case 1), one can prove that $f(x)$ has the irreducible factor $x^2 + x + 2$ and the irreducible factor
$x^4 + 2x^3 + x^2 + 1$ over $\fthree$.

\item  $(\eta(x),\eta(x+1))=(-1,-1)$: In this case one can also similarly prove that
$x^7+2x^6+x^4+x^3+2x = x(x^2 + 2x + 2)(x^4 + x^2 + 2x + 1)$
which is the canonical factorization of  $x^7+2x^6+x^4+x^3+2x$ over $\fthree$.
\end{enumerate}

Since $m$ is odd, $m \not\equiv 0 \pmod{4}$. By Lemma \ref{lem-roots-irre-poly}, (\ref{Eq-B-C2}) has the only
solution $x=1$ in $\fthreem$ if $m$ is odd. The statement that (\ref{Eq-B-C3}) has no nonzero solution in $\fthreem$
can be similarly proven for odd $m$. The desired conclusions then follow from Theorem \ref{thm-class-B}. This completes the proof.
\end{IEEEproof}

\begin{Example}
Let $m=5$ and let $\alpha$ be the generator of $\fthreem^*$ with $\alpha^5+2\alpha+1=0$.  Then the code $\mathcal{C}_{(1,e)}$ of Corollary \ref{cor-B-1} has parameters $[242, 232,4]$ and generator polynomial $x^{10}+2x^8+2x^7+2x^6+x^4+2x^2+x+2$.
\end{Example}

As an example for even $m$, we prove the following corollary.

\begin{Corollary}\label{cor-B-2}
Let $m\equiv 2 \pmod{4}$ and $e=\frac{3^m-1}{2}+10$. Then the ternary cyclic code $\mathcal{C}_{(1,e)}$ has parameters $[3^m-1,3^m-1-2m,4]$.
\end{Corollary}

\begin{IEEEproof}
Clearly $e$ is even. Hence $e\not\in C_1$.  It follows from $m\equiv 2 \pmod{4}$ that $3^m-1\equiv 0 \pmod{8}$ and $\gcd(3^m-1,5)=1$. Therefore $e\equiv 2 \pmod{4}$ and $\gcd(3^m-1,e)=\gcd(e,20)=\gcd(e,4)=2$. This leads to $|C_e|=m$ according to Lemma
\ref{lem-zero}. For $r=10$, we discuss (\ref{Eq-B-C2}) by considering the following cases:
\begin{enumerate}
\item  $(\eta(x),\eta(x+1))=(1,1)$: In this case, (\ref{Eq-B-C2}) is reduced to $2x^{10}+x^9+x+2=2(x-1)^{10}=0$
which has the only solution $x=1$.

\item  $(\eta(x),\eta(x+1))=(1,-1)$: In this case (\ref{Eq-B-C2}) is simplified to $2x(x^8+1)=0$. It is easily verified that $\gcd(x^8+1, x^{3^i}-x)=1$ for all $i \in \{1,2,3\}$. It then follows from Lemma \ref{lem-all-ir-poly} that $x^8+1$ has
no irreducible factor with degrees 1, 2 and 3. This implies that $x^8+1$ either is irreducible or has two irreducible factors of degree 4. In fact, the canonical factorization of  $x^8+1$ over $\fthree$ is
$x^8+1=(x^4 + x^2 + 2)(x^4 + 2x^2 + 2).$
Note that $x=0$ is not a solution of (\ref{Eq-B-C2}).

\item  $(\eta(x),\eta(x+1))=(-1,1)$: In this case (\ref{Eq-B-C2}) is reduced to $x^9+x+2=0$. It is straightforward to verify  that $\gcd(x^9+x-1, x^{3^i}-x)=x+1$ for all $i \in \{1,2,3\}$. It then follows from Lemma \ref{lem-all-ir-poly} that
$\frac{x^9+x-1}{x+1}$ either is irreducible or has two irreducible factors of degree 4. In fact, the canonical factorization
of $\frac{x^9+x-1}{x+1}$ over $\fthree$ is given by
$\frac{x^9+x-1}{x+1}=(x^4 + x^3 + x^2 + 1)(x^4 + x^3 + 2x^2 + 2x + 2).$
Clearly, $x=-1$ is not a solution of (\ref{Eq-B-C2}).

\item $(\eta(x),\eta(x+1))=(-1,-1)$: In this case one can similarly prove that $x^{10}+2x^9+2x$ has either two irreducible factors of degree 4 or one irreducible factor of degree 8. In fact, the canonical factorization of $x^{10}+2x^9+2x$ over $\fthree$ is given by
$x^{10}+2x^9+2x =x(x+1)(x^4 + x^2 + x + 1)(x^4 + x^3 + x^2 + 2x + 2).$
\end{enumerate}
It then follows from Lemma \ref{lem-roots-irre-poly} that (\ref{Eq-B-C2}) has the unique solution $x=1$ in $\fthreem$
if $m\equiv 2 \pmod{4}$. One can similarly prove that (\ref{Eq-B-C3}) has no solution in $\fthreem^*$  if
$m\equiv 2 \pmod{4}$.  Then the desired conclusions follow from Theorem \ref{thm-class-B}.
This completes the proof.
\end{IEEEproof}

With the same technique above, one can derive conditions on $m$ such that the ternary cyclic code $\mathcal{C}_{(1,e)}$ has parameters $[3^m-1,3^m-1-2m,4]$ for $r \in \{11, 13, 14, \cdots, 20\}$.

\begin{Example}
Let $m=6$ and let $\alpha$ be the generator of $\fthreem^*$ with $\alpha^6+2\alpha^4+\alpha^2+2\alpha+2$.
Then the code $\mathcal{C}_{(1,e)}$ of Corollary \ref{cor-B-2} has parameters $[728, 716,4]$ and generator polynomial
$x^{12}+x^{11}+2x^{10}+2x^9+2x^8+x^7+x^6+x^5+2x^2+x+2$.
\end{Example}

\subsection{The third class of optimal ternary cyclic codes with parameters  $[3^m-1,3^m-1-2m,4]$}

A class of ternary cyclic codes $\mathcal{C}_{(1,e)}$, where $e=3^m-1-2r\equiv -2r \pmod{3^m-1}$, is investigated
in this subsection. Note that $e$ is in the same 3-cyclotomic class with $r(3^{m-1}-1)$ since $3r(3^{m-1}-1)=r(3^m-3) \equiv -2r \pmod{3^m-1}$.

Clearly, $e$ is even. Hence $x=0$ and $x=-1$ are not solutions of $(-x-1)^e+x^e+1=0$. Thus Condition  C2 is satisfied if and only if $(x+1)^{-2r}+x^{-2r}+1=0$, i.e.,
\begin{equation}\label{Eq-C-C2}
 (x+1)^{2r}x^{2r}+(x+1)^{2r}+x^{2r}=0
\end{equation}
has the only solution $x=1$ in $\fthreem$. Similarly, C3 is satisfied if and only if
\begin{equation}\label{Eq-C-C3}
 (x+1)^{2r}x^{2r}+(x+1)^{2r}-x^{2r}=0
\end{equation}
has no solution $x$ in $\fthreem$.

The following theorem then follows from Theorem \ref{thm-Ding-Helleseth} and the preceding discussions.

\begin{Theorem}\label{thm-class-C}
 Let $e=r(3^{m-1}-1)$, $e \not\in C_1$ and $|C_e|=m$. Then the ternary cyclic code $\mathcal{C}_{(1,e)}$ has parameters $[3^m-1,3^m-1-2m,4]$ if (\ref{Eq-C-C2}) has the only solution $x=1$ and (\ref{Eq-C-C3}) has no solution in $\fthreem$.
\end{Theorem}

The two cases that $r=1$ and $r=3$ are covered by Theorem 6.1 in \cite{Ding-Helleseth} since $x^{3^{m-1}-1}$ is an
almost perfect nonlinear function on $\fthreem$ if $m$ is odd. For $r=2$, this is Open Problem \ref{open-one}, which
was settled before. When $r=4$ it is equivalent to $e=3^{m-2}-1$ which was discussed in Theorem 7.6 in \cite{Ding-Helleseth}. Thus, as another example, we consider the case $r=5$.

\begin{Corollary}\label{cor-C-1}
Let $m$ be odd, $m\not\equiv 0 \pmod{3}$ and $e=5(3^{m-1}-1) \equiv 2(3^{m-1}-2) \pmod{3^m-1}$. Then the ternary cyclic code $\mathcal{C}_{(1,e)}$ has parameters $[3^m-1,3^m-1-2m,4]$.
\end{Corollary}

\begin{IEEEproof}
Since $e$ is even, $e \not\in C_1$. It is easily seen that $\gcd(e,3^m-1)=2$. Then it follows from Lemma \ref{lem-zero}
that $|C_e|=m$.
When $r=5$,  (\ref{Eq-C-C2}) is reduced to $f(x)=(x+1)^{10}x^{10}+(x+1)^{10}+x^{10}=x^{20}+x^{19}+x^{11}+x^9+x+1$. By Lemma \ref{lem-Division-poly}, one can derive that $\gcd(f(x), x^3-x)=x-1$, $\gcd(f(x), x^{3^3}-x)=x^7-x^6-x^5+x^2+x-1$  and $\gcd(f(x), x^{3^5}-x)=x-1$.  This together with Lemma \ref{lem-all-ir-poly} implies that $f(x)$ has two cubic irreducible factors and has no irreducible factor of degree $5$. Hence, one can claim that $f(x)$ has no other irreducible factor with odd degree due to the facts $(x-1)^2|f(x)$ and $\deg(f(x))=20$. In fact, the canonical factorization of $f(x)$ over $\fthree$ is given by $f(x)=(x-1)^2(x^3 + x^2 + x + 2)(x^3 + 2x^2 + 2x + 2)(x^6 + x^5 + 2x^3 + x^2 + 2x + 1)(x^6 + 2x^5 + x^4 + 2x^3 + x + 1)$.
Then by Lemma \ref{lem-roots-irre-poly}, $f(x)=0$ has the only solution $x=1$ in $\fthreem$ if $m$ is odd and $m\not\equiv 0 \pmod{3}$.

Similarly, one can prove that (\ref{Eq-C-C3}) has no solution $x$ in $\fthreem$. When $r=5$,  (\ref{Eq-C-C3}) is simplified to $g(x)=x^{20}+x^{19}+x^{11}+x^{10}+x^9+x+1=0$.
It is straightforward to check that
\begin{eqnarray*}
&& \gcd(g(x), x^3-x)=\gcd(g(x), x^{3^2}-x)=1, \\
&& \gcd(g(x), x^{3^3}-x)=x^6+x^5-x^4-x^2+x+1, \\
&& \gcd(g(x), x^{3^4}-x)=x^8-x^7+x^6+x^5+x^4+x^3+x^2-x+1.
\end{eqnarray*}
It then follows from Lemma \ref{lem-all-ir-poly} that $g(x)$ has two cubic irreducible factors and two irreducible factors of degree $4$. This implies that $g(x)$ has no other irreducible factors with odd degree.  In fact, the canonical factorization of $g(x)$ over $\fthree$ is given by $g(x)=(x^3 + 2x + 2)(x^3 + x^2 + 2)(x^4 + x^3 + x^2 + 2x + 2)(x^4 + x^3 + 2x^2 + 2x + 2)(x^6 + x^5 + x^4 + x^3 + x^2 + x + 1)$.
Thus, (\ref{Eq-C-C3}) has no solution $x$ in $\fthreem$ if $m\not\equiv 0 \pmod{3}$ and $m\not\equiv 0 \pmod{4}$.  Then the desired conclusions follow from Theorem \ref{thm-class-C}.   This completes the proof.
\end{IEEEproof}

\begin{Example}
Let $m=5$ and let $\alpha$ be the generator of $\fthreem^*$ with $\alpha^5+2\alpha+1=0$.  Then the code $\mathcal{C}_{(1,e)}$ of Corollary \ref{cor-C-1} has parameters $[242, 232,4]$ and generator polynomial $x^{10}+x^9+x^7+x^6+2x^5+x^4+2x^3+2x^2+2$.
\end{Example}

\subsection{Two more classes of optimal ternary cyclic codes with parameters  $[3^m-1,3^m-1-2m,4]$}

In this subsection we treat small integers $e$ such that  the ternary cyclic code $\mathcal{C}_{(1,e)}$ is optimal.
Specifically, we consider the ternary cyclic codes $\mathcal{C}_{(1,e)}$, where $e=2r$ and $1\leq r\leq 10$.
Most of them were studied in \cite{Carlet-Ding-Yuan} and \cite{Ding-Helleseth}:
\begin{enumerate}
    \item When $e \in \{2, 6, 18\}$, the code is covered by Theorem 5.2 in \cite{Ding-Helleseth} since they lie in the same $3$-cyclotomic coset and $x^2$ is a planar function over $\fthreem$.

    \item When $e \in \{4,12\}$, the code is covered by Theorem 5.2 in \cite{Ding-Helleseth} since 4 and 12 are in the same $3$-cyclotomic coset and $x^{3^h+1}$ is a planar function over $\fthreem$ if $m/\gcd(m,h)$ is odd.

    \item When $e=8$, the code is covered by  Theorem 7.6 in \cite{Ding-Helleseth} since $e$ can be written as $e=3^2-1$.

    \item When $e=10$, the code is covered by Theorem 5.2 in \cite{Ding-Helleseth} since $x^{3^h+1}$ is a planar function over $\fthreem$ if $m/\gcd(m,h)$ is odd.

    \item When $e=14$, the code is covered by Theorem 5.2 in \cite{Ding-Helleseth} since $x^{\frac{3^h+1}{2}}$ is a planar function over $\fthreem$ if $\gcd(m,h)=1$ and $h$ is odd.
  \end{enumerate}
Thus, the remaining cases are $e=16$ and $e=20$. In what follows, we investigate the two codes $\mathcal{C}_{(1,16)}$ and $\mathcal{C}_{(1,20)}$

\begin{Corollary}\label{cor-e=16}
Let $m$ be odd and $m\not\equiv 0 \pmod{3}$. Then the ternary cyclic code $\mathcal{C}_{(1,16)}$ has parameters $[3^m-1,3^m-1-2m,4]$.
\end{Corollary}

\begin{IEEEproof}
Clearly $e=16 \not\in C_1$. Note that $3^m-1\equiv 2 \pmod{4}$ as $m$ is odd. We have obviously that $\gcd(16,3^m-1)=2$.
It then follows from Lemma \ref{lem-zero} that $|C_e|=m$. The condition C2 is met for $e=16$ if and only if $(x+1)^{16}+x^{16}+1=0$ has the only solution $x=1$ in $\fthreem$. Note that
$$
\frac{(x+1)^{16}+x^{16}+1}{(x-1)^4} =-(x^{12}-x^9-x^8-x^7-x^5-x^4-x^3+1).
$$
It suffices to prove that
  \begin{equation}\label{Eq-e=16-C2}
f(x):=  x^{12}-x^9-x^8-x^7-x^5-x^4-x^3+1=0
  \end{equation}
has no solution in $\fthreem$.
It is straightforward to obtain that $\gcd(f(x),x^3-x)=1$,  $\gcd(f(x),x^{3^2}-x)=x^6+x^4+x^2+1$ and $\gcd(f(x),x^{3^3}-x)=x^6-x^4-x^3-x^2+1$. It then follows from Lemma \ref{lem-all-ir-poly} that $f(x)$ has three quadratic irreducible factors and two cubic irreducible factors. In fact, the canonical factorization of $f(x)$ over $\fthree$ is given by $f(x)=(x^2 + 1)(x^2 + x + 2) (x^2 + 2x + 2) (x^3 + x^2 + x + 2) (x^3 + 2x^2 + 2x + 2)$.
Hence (\ref{Eq-e=16-C2}) has no solution in $\fthreem$ if and only if $m$ is odd and $m\not\equiv 0 \pmod{3}$.

Condition C3 can be similarly treated.  Condition C3 is satisfied for $e=16$ if and only if $(x+1)^{16}-x^{16}-1=0$ has the only solution $x=0$ in $\fthreem$. Note that
$$
(x+1)^{16}-x^{16}-1=x(x^{14}-x^{12}-x^{11}+x^9+x^8+x^6+x^5-x^3-x^2+1).
$$
We need to prove that
\begin{equation}\label{Eq-e=16-C3}
g(x):=  x^{14}-x^{12}-x^{11}+x^9+x^8+x^6+x^5-x^3-x^2+1=0
\end{equation}
has no solution in $\fthreem$. One can verify that
$\gcd(g(x),x^3-x)=1$, $\gcd(g(x),x^{3^2}-x)=1$ and $\gcd(g(x),x^{3^3}-x)=x^6+x^5-x^4-x^2+x+1$.
It then follows from Lemma \ref{lem-all-ir-poly} that $g(x)$ has two cubic irreducible factors and $g(x)/\gcd(g(x),x^{3^3}-x)$ either is irreducible or has two irreducible factors of degree $4$. In fact, the canonical factorization of $g(x)$ over $\fthree$
is given by $g(x)=(x^3 + 2x + 2)(x^3 + x^2 + 2)(x^8 + 2x^7 + x^6 + 2x^4 + x^2 + 2x + 1)$.
Thus (\ref{Eq-e=16-C3}) has no solution in $\fthreem$ if $m$ is odd and $m\not\equiv 0 \pmod{3}$. Then the desired conclusions follow from Theorem \ref{thm-Ding-Helleseth}.  This completes the proof.
\end{IEEEproof}

\begin{Example}
Let $m=5$ and let $\alpha$ be the generator of $\fthreem^*$ with $\alpha^5+2\alpha+1=0$.  Then the code $\mathcal{C}_{(1,e)}$ of Corollary \ref{cor-e=16} has parameters $[242, 232,4]$ and generator polynomial $x^{10}+2x^9+x^8+x^7+x^5+x^4+2x+2$.
\end{Example}

   \begin{Corollary}\label{cor-e=20}
   Let $m$ be odd. Then the ternary cyclic code $\mathcal{C}_{(1,20)}$ has parameters $[3^m-1,3^m-1-2m,4]$.
  \end{Corollary}

\begin{IEEEproof}
Since $e=20$ is even, $e \not\in C_1$.  We have that $\gcd(3^m-1,20)=\gcd(3^m-1, 4)=2$ as $m$ is odd. It then follows
from Lemma \ref{lem-zero} that $|C_e|=|C_{20}|=m$. Conditions C2 and C3 are met if and only if both $f(x)=[(x+1)^{20}+x^{20}+1]/(x-1)^2=2(x^{18}+x^{16}-x^{15}+x^{13}-x^{12}+x^{10}+x^8-x^6+x^5-x^3+x^2+1)$
  and $g(x)=[(x+1)^{20}-x^{20}-1]/x=2(x^{18}-x^{17}+x^{10}-x^9+x^8-x+1)$ have no solution in $\fthreem$. By Lemma \ref{lem-roots-irre-poly}, it is sufficient to prove that both $f(x)$ and $g(x)$ have no irreducible factors with odd degree.

Applying Lemma \ref{lem-Division-poly}, one obtains that
$$
\gcd(f(x), x^{3^2}-x)=1, \gcd(f(x), x^{3^4}-x)=1, \gcd(f(x), x^{3^6}-x)=2f(x),
$$
i.e., $f(x)$ has three irreducible factors of degree $6$. Then $f(x)=0$ has no solution in $\fthreem$ if and only if $m\not\equiv 0 \pmod{6}$ due to Lemma \ref{lem-roots-irre-poly}. For $g(x)$, one has that  $\gcd(g(x), x^{3^2}-x)=x^2+1$ and $\gcd(g(x), x^{3^4}-x)=x^6+x^5-x^4-x^3-x^2+x+1$. It then follows from
Lemma \ref{lem-roots-irre-poly} that $g(x)$ has the irreducible factor $x^2+1$ and an irreducible factor of degree $4$. Moreover, Lemma \ref{lem-all-ir-poly} implies that $g(x)/\gcd(g(x), x^{3^4}-x)$ whose degree equals to $12$ has no irreducible factors with odd degree since $\gcd(g(x), x^{3^3}-x)=1$ and $\gcd(g(x), x^{3^5}-x)=1$. Thus, by  Lemma \ref{lem-roots-irre-poly}, $g(x)=0$ has no solution in $\fthreem$ if $m$ is odd. Then the desired conclusions follow from Theorem \ref{thm-Ding-Helleseth}.  This completes the proof.
\end{IEEEproof}

\begin{Remark}
The value $e=20$ is a special case of Open Problem 7.5 in \cite{Ding-Helleseth}.
\end{Remark}

\begin{Example}
Let $m=5$ and let $\alpha$ be the generator of $\fthreem^*$ with $\alpha^5+2\alpha+1=0$.  Then the code $\mathcal{C}_{(1,e)}$ of Corollary \ref{cor-e=20} has parameters $[242, 232,4]$ and generator polynomial $x^{10}+2x^8+x^7+x^4+x^3+2x+2$.
\end{Example}

It should be noticed that some larger values of $r$ for $e=\frac{3^m-1}{2}\pm r$, $e=r(3^{m-1}-1)$ and $e=2r$ such that  $\mathcal{C}_{(1,e)}$ has parameters $[3^m-1,3^m-1-2m,4]$ can also be obtained with the same techniques by virtue of the division algorithm given in Lemma \ref{lem-Division-poly}. Moreover, in each of the eight remaining open problems in \cite{Ding-Helleseth}, the cases of $h=0,1,2, 3$ and $h=m-1, m-2, m-3$ can also be settled. For the general case, new techniques are required.

\section{New optimal ternary cyclic codes with parameters $[3^m-1,3^m-2m-2, 5]$ }

 Throughout this section, let $m>1$ be an integer, $s=\frac{3^m-1}{2}$, $\alpha$ be a generator of $\fthreem^*$ and $\mathcal{C}_{(1,e,s)}$ be the cyclic code with generator polynomial $(x+1)m_{\alpha}(x)m_{\alpha^e}(x)$, where $m_{\alpha^i}(x)$ denotes the minimal polynomial of $\alpha^i$ over $\fthree$.

 The code $\mathcal{C}_{(1,e,s)}$ is a $[3^m-1, 3^m-2m-2]$ cyclic code if the size of $C_e$ is equal to $m$, i.e., $|C_e|=m$. A tight upper bound on the minimum distance of $\mathcal{C}_{(1,e,s)}$ can be derived from the following bound on linear codes.

\begin{Lemma}\label{lem-RGSS-bound}
(\rm \cite[Lemma 6]{Rouayheb-GSS})
Let $A_q(n,d)$ be the maximum number of codewords of a $q$-ary code with length $n$ and Hamming distance at least $d$. If $q\geq 3$, $t=n-d+1$ and $r=\lfloor\min\{\frac{n-t}{2},\frac{t-1}{q-2}\}\rfloor$, then
     \[A_q(n,d)\leq\frac{q^{t+2r}}{\sum_{i=0}^r\binom {t+2r}{i}(q-1)^i}.\]
\end{Lemma}

 \begin{Theorem} \label{thm-d<=5}
For any given $e$ with $|C_e|=m$, the minimum distance of $\mathcal{C}_{(1,e,s)}$ satisfies $d \leq 5$.
\end{Theorem}

\begin{IEEEproof}
Clearly, $\mathcal{C}_{(1,e,s)}$ has length $3^m-1$ and dimension $3^m-2m-2$ if $|C_e|=m$. It follows from the Sphere Packing bound that the minimum distance $d$ of $\mathcal{C}_{(1,e,s)}$ satisfies $d\leq 6$.  It then suffices to show there is no ternary code with  parameters $[3^m-1, 3^m-2m-2, 6]$.

Assume that there exists a ternary code with parameters $[3^m-1, 3^m-2m-2,6]$. Then applying Lemma \ref{lem-RGSS-bound} to this code, we have $q=3$, $n=3^m-1$, $t=n-d+1=3^m-6$, $r=2$, $t+2r=3^m-2$, $\sum_{i=0}^r\binom {t+2r}{i}(q-1)^i=1+2(3^m-2)^2$, and
   \[3^{3^m-2m-2}\leq A_3(3^m-1,6)\leq\frac{3^{3^m-2}}{1+2(3^m-2)^2}\]
which implies $1+2(3^m-2)^2\leq 3^{2m}$, i.e., $(3^m-4)^2\leq 7$, a contradiction is obtained if $m>1$. This completes the proof.
\end{IEEEproof}

Theorem \ref{thm-d<=5} indicates that $\mathcal{C}_{(1,e,s)}$ is optimal if it has parameters $[3^m-1, 3^m-2m-2,5]$. Thus, for any given $e$, in order to obtain optimal ternary cyclic code $\mathcal{C}_{(1,e,s)}$ with parameters $[3^m-1, 3^m-2m-2,5]$, we need to show that $\mathcal{C}_{(1,e,s)}$ has no codeword of Hamming weights $\omega \in \{1,2,3,4\}$. By the definition of $\mathcal{C}_{(1,e,s)}$, it has a codeword of Hamming weight $\omega$ if and only if there exist $\omega$ nonzero elements $c_1,c_2,\cdots,c_{\omega}$ in $\fthree$ and $\omega$ nonzero distinct elements $x_1,x_2,\cdots,x_{\omega}$ in $\fthreem$ such that
   \begin{equation}\label{Eq-(x+1)-w}
    \left\{\begin{array}{lll}
       c_1x_1+c_2x_2+\cdots+c_{\omega}x_{\omega}=0
    \\ c_1x_1^e+c_2x_2^e+\cdots+c_{\omega}x_{\omega}^e=0
    \\ c_1x_1^s+c_2x_2^s+\cdots+c_{\omega}x_{\omega}^s=0.
  \end{array} \right.
 \end{equation}
Clearly, (\ref{Eq-(x+1)-w}) cannot hold for $\omega=1$. If $\omega=2$, then one has $c_1=c_2$ since $x_1\ne x_2$. This implies $0=c_1x_1^e+c_2x_2^e=c_1(x_1^e+(-x_1)^e)$ and $0=c_1x_1^s+c_2x_2^s=c_1(x_1^s+(-x_1)^s)$, i.e., $\omega \ne 2$ if and only if either $e$ is even or $s$ is even.  Note that $s=\frac{3^m-1}{2}$ is even only if $m$ is even.

To consider the codewords in $\mathcal{C}_{(1,e,s)}$ with Hamming weights $\omega=3$ and $\omega=4$, it is more convenient to write (\ref{Eq-(x+1)-w}) as
\begin{equation}\label{Eq-(x+1)-w=1+}
    \left\{\begin{array}{lll}
       1+\frac{c_2x_2}{c_1x_1}+\cdots+\frac{c_{\omega}x_{\omega}}{c_1x_1}=0
    \\ 1+\frac{c_2x_2^e}{c_1x_1^e}+\cdots+\frac{c_{\omega}x_{\omega}^e}{c_1x_1^e}=0
    \\ 1+\frac{c_2 x_2^s}{c_1x_1^s}+\cdots+\frac{c_{\omega}x_{\omega}^s}{c_1 x_1^s}=0.
  \end{array} \right.
 \end{equation}

In the subsequent subsections, by discussing the solutions of (\ref{Eq-(x+1)-w=1+}) for $\omega=3$ and $\omega=4$ for a given $e$, new optimal ternary cyclic codes $\mathcal{C}_{(1,e,s)}$ with parameters $[3^m-1,3^m-2m-2, 5]$ will be obtained.

\subsection{New optimal double-error-correcting ternary cyclic codes for even $m$}

In this subsection, new optimal ternary cyclic codes $\mathcal{C}_{(1,e,s)}$ with parameters $[3^m-1,3^m-2m-2, 5]$ will be obtained from the exponent $e$ of the form
\begin{eqnarray}\label{eqn-edef}
e=\frac{3^m-1}{2}+r, \ \ 1\leq r \leq 3^m-2,
\end{eqnarray}
where $x^r$ is PN over $\fthreem$.

By definition, a function from $\fthreem$ to itself is PN if and only if
\begin{eqnarray}\label{Eq-PN-property}
\left\{ \begin{array}{l}
x-y = a \\
f(x)-f(y)=b
\end{array}
\right.
\end{eqnarray}
has a unique solution $(x, y) \in \fthreem \times \fthreem$ for each $(a, b) \in \fthreem^* \times \fthreem$.

The following is a list of known PN monomials over $\fthreem$:
\begin{itemize}
     \item $f(x)=x^2$;
     \item $f(x)=x^{3^h+1}$, where $m/\gcd(m,h)$ is odd \cite{Dembowski-Ostrom};
     \item $f(x)=x^{\frac{3^h+1}{2}}$, where $\gcd(m,h)=1$ and $h$ is odd \cite{Coulter-Matthews}.
\end{itemize}

It is known that $r$ must be even if $x^r$ is PN over $\fthreem$. This fact will be frequently used in subsequent proofs.
If $m$ is odd and $x^r$ is PN over $\fthreem$, then the integer $e$ of (\ref{eqn-edef}) must be odd as $\frac{3^m-1}{2}$
is odd. In this case the minimum distance $d$ of $\mathcal{C}_{(1,e,s)}$ is $2$.  Hence throughout this section we assume
that $m$ is even and $x^r$ is PN over $\fthreem$.  Let $e=\frac{3^m-1}{2}+r$.  Under these assumptions we will prove
that $\mathcal{C}_{(1,e,s)}$ is optimal and  has parameters $[3^m-1,3^m-2m-2, 5]$. To this end, we need to prove that  $\mathcal{C}_{(1,e,s)}$ has no codeword of Hamming weights 3 and 4.

For simplicity, from now on, let $\eta$ denote the quadratic character on $\fthreem$ which is defined by $\eta(x)=1$ if $x$ is a nonzero square in $\fthreem$ and $\eta(x)=-1$ if $x$ is a nonzero nonsquare in $\fthreem$.

\begin{Lemma}\label{lem-1st-d-not=3}
Let $m$ be even, $s=\frac{3^m-1}{2}$ and $e=\frac{3^m-1}{2}+r$, where $1\leq r\leq 3^m-2$. Then $\mathcal{C}_{(1,e,s)}$ has no codeword of Hamming weight 3 if $f(x)=x^r$ is PN over $\fthreem$.
\end{Lemma}

\begin{IEEEproof}
$\mathcal{C}_{(1,e,s)}$ has no codeword of Hamming weight $\omega=3$ if and only if (\ref{Eq-(x+1)-w=1+}) has no solution over $\fthreem$ for $\omega=3$. Let $x=x_2/x_1$ and $y=x_3/x_1$, then $x,y\ne 0,1$, $x\ne y$ and  (\ref{Eq-(x+1)-w=1+}) becomes
 \begin{equation}\label{Eq-1st-d-not=3-proof}
 \left\{\begin{array}{lll}
       1+\frac{c_2}{c_1}x+\frac{c_3}{c_1}y=0
    \\ 1+\frac{c_2}{c_1}\eta(x)x^r+\frac{c_3}{c_1}\eta(y)y^r=0
    \\1+\frac{c_2}{c_1}\eta(x)+\frac{c_3}{c_1}\eta(y)=0
  \end{array} \right.
  \end{equation}
  since $x^e=\eta(x)x^r$ and $y^e=\eta(y)y^r$. Due to symmetry, we only need to consider (\ref{Eq-1st-d-not=3-proof}) for the following two cases:
  \begin{enumerate}
    \item $c_1=c_2=c_3=1$: In this case, by the third equation in (\ref{Eq-1st-d-not=3-proof}), one has $1+\eta(x)+\eta(y)=0$. This leads to $\eta(x)=\eta(y)=1$. Then (\ref{Eq-1st-d-not=3-proof}) is reduced to $1+x+y=0$ and $1+x^r+y^r=0$.
    \item $c_1=c_2=1$, $c_3=-1$: Similarly, by (\ref{Eq-1st-d-not=3-proof}), one has $1+\eta(x)-\eta(y)=0$, i.e., $\eta(x)=1$ and $\eta(y)=-1$.  Thus (\ref{Eq-1st-d-not=3-proof}) is reduced to $1+x-y=0$ and $1+x^r+y^r=0$.
  \end{enumerate}
 Assume that $(x,y)$ is a solution of (\ref{Eq-1st-d-not=3-proof}) with $x,y\ne 0,1$ and $x\ne y$. Then 1) and 2) imply that
  \[x-1=1-(\pm y), \; x^r-1^r=1^r-(\pm y)^r\]
where we used the fact that $r$ is even (because $x^r$ is PN over $\fthreem$). Furthermore, since $f(x)=x^r$ is PN over $\fthreem$, by (\ref{Eq-PN-property}),  the equations above hold if and only if $(x,1)=(1,\pm y)$, a contradiction with $x\ne 1$. Thus, (\ref{Eq-(x+1)-w=1+}) has no solution for $\omega=3$. This completes the proof.
\end{IEEEproof}

With the same techniques, we can also prove $\mathcal{C}_{(1,e,s)}$ has no codeword of Hamming weight 4 if $e=\frac{3^m-1}{2}+r$ and $f(x)=x^r$ is PN over $\fthreem$.

\begin{Lemma}\label{lem-1st-d-not=4}
Let $m$ be even, $s=\frac{3^m-1}{2}$ and $e=\frac{3^m-1}{2}+r$, where $1\leq r\leq 3^m-2$. Then $\mathcal{C}_{(1,e,s)}$ has no codeword of Hamming weight 4 if $f(x)=x^r$ is PN over $\fthreem$.
\end{Lemma}

 \begin{IEEEproof} $\mathcal{C}_{(1,e,s)}$ has no codeword of Hamming weight $\omega=4$ if and only if (\ref{Eq-(x+1)-w=1+}) has no solution over $\fthreem$ for $\omega=4$. Let $x=x_2/x_1$, $y=x_3/x_1$ and $z=x_4/x_1$, then $x,y,z\ne 0,1$ are pairwise distinct and  (\ref{Eq-(x+1)-w=1+}) becomes
 \begin{equation}\label{Eq-1st-d-not=4-proof}
 \left\{\begin{array}{lll}
       1+\frac{c_2}{c_1}x+\frac{c_3}{c_1}y+\frac{c_4}{c_1}z=0
    \\ 1+\frac{c_2}{c_1}\eta(x)x^r+\frac{c_3}{c_1}\eta(y)y^r+\frac{c_4}{c_1}\eta(z)z^r=0
    \\1+\frac{c_2}{c_1}\eta(x)+\frac{c_3}{c_1}\eta(y)+\frac{c_4}{c_1}\eta(z)=0
  \end{array} \right.
  \end{equation}
 since $x^e=\eta(x)x^r$, $y^e=\eta(y)y^r$ and $z^e=\eta(z)z^r$ if $e=\frac{3^m-1}{2}+r$. Due to symmetry, we only need to consider (\ref{Eq-1st-d-not=4-proof}) for the following three cases:
  \begin{enumerate}
    \item $c_1=c_2=c_3=c_4=1$: By the third equation in (\ref{Eq-1st-d-not=4-proof}), i.e., $1+\eta(x)+\eta(y)+\eta(z)=0$, without loss of generality, one can assume $\eta(x)=1$ and $\eta(y)=\eta(z)=-1$. Then (\ref{Eq-1st-d-not=4-proof}) is reduced to $1+x+y+z=0$ and $1+x^r-y^r-z^r=0$.
    \item $c_1=c_2=c_3=1$, $c_4=-1$: In this case, one has $1+\eta(x)+\eta(y)-\eta(z)=0$ according to (\ref{Eq-1st-d-not=4-proof}).  Then there are two cases to be considered: (I) If $\eta(z)=-1$, then $\eta(x)+\eta(y)=1$ which implies $\eta(x)=\eta(y)=-1$; (II) If $\eta(z)=1$, then $\eta(x)+\eta(y)=0$ and one can assume $\eta(x)=1$ and $\eta(y)=-1$ due to the symmetry of $x$ and $y$. Then, (\ref{Eq-1st-d-not=4-proof}) can be reduced to $1+x+y-z=0$ and $1-x^r-y^r+z^r=0$ or  $1+x+y-z=0$ and $1+x^r-y^r-z^r=0$.
    \item $c_1=c_2=1$, $c_3=c_4=-1$: Similar as case 2), by (\ref{Eq-1st-d-not=4-proof}) one has $1+\eta(x)-\eta(y)-\eta(z)=0$. (I) If $\eta(x)=1$, then $\eta(x)+\eta(y)=-1$. This leads to $\eta(y)=\eta(z)=1$; (II) If $\eta(x)=-1$, then $\eta(y)+\eta(z)=0$ and one can assume $\eta(y)=1$ and $\eta(z)=-1$ due to the symmetry of $y$ and $z$. Therefore, (\ref{Eq-1st-d-not=4-proof}) can be reduced to $1+x-y-z=0$ and $1+x^r-y^r-z^r=0$ or  $1+x-y-z=0$ and $1-x^r-y^r+z^r=0$.
   \end{enumerate}
  Thus there are totally five cases to be considered. However, it should be noticed that there are exactly two ``$-1$" and two ``$1$" in the multi-set $\{1,\frac{c_2}{c_1}\eta(x),\frac{c_3}{c_1}\eta(y), \frac{c_4}{c_1}\eta(z)\}$ for each of the five cases. This makes the proof for each case quite similar by using properties of PN functions. Hence we only prove  that (\ref{Eq-1st-d-not=4-proof}) has no solution over $\fthreem$ for the first case.  Assume that $(x,y,z)$ is a solution over $\fthreem$ of (\ref{Eq-1st-d-not=4-proof}) with the conditions $c_1=c_2=c_3=c_4=1$, $\eta(x)=1$ and $\eta(y)=\eta(z)=-1$. Then (\ref{Eq-1st-d-not=4-proof}) can be rewritten as
   \[x-(-y)=(-z)-1,\; \ x^r-(-y)^r=(-z)^r-1^r\]
   since $r$ is even.  However the above equalities hold if and only if $(x,-y)=(-z,1)$ according to (\ref{Eq-PN-property}) since $x^r$ is PN over $\fthreem$. This contradicts with $\eta(y)=-1$ for the first case since $\eta(-1)=1$ for even $m$.  This completes the proof.
   \end{IEEEproof}

 For even $m$, new classes of optimal ternary cyclic codes with parameters $[3^m-1,3^m-2m-2, 5]$ can be obtained
 from the known PN monomials over $\fthreem$ and are described in the following theorem.

\begin{Theorem} \label{thm-even-m-PN}
    Let $m$ be even and $s=\frac{3^m-1}{2}$. Then the ternary cyclic code $\mathcal{C}_{(1,e,s)}$ is optimal and has parameters $[3^m-1,3^m-2m-2, 5]$ if
    \begin{itemize}
      \item $e=\frac{3^m-1}{2}+2$; or
      \item $e=\frac{3^m-1}{2}+3^h+1$, where $m/\gcd(m,h)$ is odd; or
      \item $e=\frac{3^m-1}{2}+\frac{3^h+1}{2}$, where $\gcd(m,h)=1$ and $h$ is odd.
    \end{itemize}
  \end{Theorem}

\begin{IEEEproof}
For each class of $e$, one can derive that $\gcd(e,3^m-1)=2.$ By Lemma \ref{lem-zero}, we have that $|C_e|=m$. It can be easily verified that $C_1 \cap C_e = \emptyset$. Then the desired conclusions follow from Lemmas \ref{lem-1st-d-not=3}--\ref{lem-1st-d-not=4}, Theorem \ref{thm-d<=5} and the fact that the three monomials $x^2$, $x^{3^h+1}$
and $x^{(3^h+1)/2}$ are PN over $\fthreem$ under the conditions described in this theorem. This completes the proof.
\end{IEEEproof}

  \begin{Example} Two examples of the codes of Theorem \ref{thm-even-m-PN} are the following:
    \begin{enumerate}
      \item Let $m=4$, $s=\frac{3^m-1}{2}=40$ and $\alpha$ be a generator of $\fthreem^*$ with $\alpha^4+2\alpha^3+2=0$. If $e=\frac{3^m-1}{2}+2=42$, then the generator polynomial of $\mathcal{C}_{(1,42,40)}$ is $x^9+x^8+2x^6+2x^5+x^4+2x^2+2x+2$ and $\mathcal{C}_{(1,42,40)}$ has parameters $[80,71,5]$.
      \item Let $m=6$, $s=\frac{3^m-1}{2}=364$ and $\alpha$ be a generator of $\fthreem^*$ with $\alpha^6+2\alpha^4+\alpha^2+2\alpha+2=0$. If $e=\frac{3^m-1}{2}+3^2+1=374$, then the generator polynomial of $\mathcal{C}_{(1,374,364)}$ is $x^{13}+2x^{12}+x^{10}+x^9+2x^7+2x^6+x^5+2x^3+2$ and $\mathcal{C}_{(1,374,364)}$ has parameters $[728,715,5]$.
    \end{enumerate}
According to Theorem \ref{thm-even-m-PN}, both $\mathcal{C}_{(1,42,40)}$ and $\mathcal{C}_{(1,374,364)}$ are optimal.
\end{Example}

Another class of optimal ternary cyclic code $\mathcal{C}_{(1,e,s)}$ with parameters $[3^m-1,3^m-2m-2, 5]$ for even $m$
are described in the following theorem.

\begin{Theorem} \label{thm-even-m-e=2}
Let $m$ be even and let $s=\frac{3^m-1}{2}$. Then the ternary cyclic code $\mathcal{C}_{(1,2,s)}$ is optimal and  has parameters $[3^m-1,3^m-2m-2, 5]$.
\end{Theorem}

\begin{IEEEproof}
It follows from Lemma \ref{lem-zero} that $|C_2|=m$. Clearly, we have that $C_1 \cap C_2=\emptyset$. Hence the dimension of $\mathcal{C}_{(1,2,s)}$ is
equal to $3^m-2m-2$.  By the BCH bound, the minimum distance of $\mathcal{C}_{(1,2)}$ is at least $4$. Then to prove this theorem, we need only to prove that (\ref{Eq-(x+1)-w=1+}) has no solution over $\fthreem$ for $e=2$ and $\omega=4$ due to the fact that $\mathcal{C}_{(1,2,s)}\subset \mathcal{C}_{(1,2)}$ and Theorem \ref{thm-d<=5}. Because of symmetry, we discuss (\ref{Eq-(x+1)-w=1+}) by distinguishing among the following cases:
\begin{enumerate}
 \item $c_1=c_2=c_3=c_4=1$: Without loss of generality, we can assume $\eta(x)=1$ and $\eta(y)=\eta(z)=-1$ by the third equation in (\ref{Eq-(x+1)-w=1+}). On the other hand, the first two equations in (\ref{Eq-(x+1)-w=1+}) imply that  $1+x^2+y^2+(-1-x-y)^2=0$, i.e., $x^2+(y+1)x+y^2+y+1=0$. For any fixed $y$, the discriminant of this quadratic equation with unknown $x$ is given by $\Delta=(y+1)^2-4(y^2+y+1)=y$ which is a nonsquare since $\eta(y)=-1$. Hence, (\ref{Eq-(x+1)-w=1+}) has no solution over $\fthreem$ for this case.
\item $c_1=c_2=c_3=1$, $c_4=-1$: In this case, by the first two equations in (\ref{Eq-(x+1)-w=1+}) one has $1+x^2+y^2-(1+x+y)^2=0$, i.e., $x+y+xy=0$. This implies $y=-\frac{x}{x+1}$ and $\frac{1}{x+1}=-\frac{y}{x}$. Then $\eta(z)=\eta(1+x+y)=\eta(\frac{(x-1)^2}{x+1})=\eta(\frac{1}{x+1})=\eta(-\frac{y}{x})=\eta(-xy)=\eta(xy)$ since $\eta(-1)=1$ if $m$ is even. This leads to  $0=1+\eta(x)+\eta(y)-\eta(z)=1+\eta(x)+\eta(y)-\eta(xy)=(\eta(x)-1)(1-\eta(y))-1\in\{1,-1\}$ since $\eta(x),\eta(y)\in\{1,-1\}$, a contradiction. Thus, (\ref{Eq-(x+1)-w=1+}) has no solution over $\fthreem$.
\item $c_1=c_2=1$, $c_3=c_4=-1$: The first two equations in (\ref{Eq-(x+1)-w=1+}) imply that $1+x^2-y^2-(1+x-y)^2=0$ for this case, i.e, $y^2-y-xy+x=(y-x)(y-1)=0$ which is impossible since $x\ne y$ and $y\ne 1$.
\end{enumerate}
 Thus, (\ref{Eq-(x+1)-w=1+}) has no solution over $\fthreem$ for $e=2$ and $\omega=4$.   This completes the proof.
\end{IEEEproof}

It should be noted that the codes of Theorem \ref{thm-even-m-e=2} are not BCH codes.

\begin{Example} The following are two examples of the codes of Theorem \ref{thm-even-m-e=2}.
\begin{enumerate}
\item Let $m=4$, $s=\frac{3^m-1}{2}=40$ and $\alpha$ be a generator of $\fthreem^*$ with $\alpha^4+2\alpha^3+2=0$. Then the generator polynomial of $\mathcal{C}_{(1,2,40)}$ is $x^9+2x^8+x^6+2x^5+2x^3+2x^2+2x+2$ and $\mathcal{C}_{(1,2,40)}$ has parameters $[80,71,5]$.
\item Let $m=6$, $s=\frac{3^m-1}{2}=364$ and $\alpha$ be a generator of $\fthreem^*$ with $\alpha^6+2\alpha^4+\alpha^2+2\alpha+2=0$. Then the generator polynomial of $\mathcal{C}_{(1,2,364)}$ is $x^{13}+2x^{12}+2x^9+2x^8+x^5+x^4+2x^3+x+2$ and $\mathcal{C}_{(1,2,364)}$ has parameters $[728,715,5]$.
\end{enumerate}
According to Theorem \ref{thm-even-m-e=2}, both $\mathcal{C}_{(1,2,40)}$ and $\mathcal{C}_{(1,2,364)}$ are optimal.
\end{Example}

\subsection{New optimal double-error-correcting ternary cyclic codes for odd $m$}

In this subsection, new optimal ternary cyclic codes $\mathcal{C}_{(1,e,s)}$ with parameters $[3^m-1,3^m-2m-2, 5]$ for odd
$m$ are investigated, where $e$ is even and satisfies
\begin{equation}\label{Eq-def-e}
e\cdot r\equiv 2\cdot 3^{\tau} \pmod{3^m-1}
\end{equation}
for some $1\leq r\leq 3^m-2$ and $0\leq \tau\leq m-1$.

To prove that $\mathcal{C}_{(1,e,s)}$ has minimum distance $d=5$, we need to show that $\mathcal{C}_{(1,e,s)}$ has no codeword of Hamming weights 3 and 4.

\begin{Lemma}\label{lem-odd-m-w-not=3}
Let $m$ be odd and let $s=\frac{3^m-1}{2}$. Let $e$ and $r$ be even positive integers satisfying (\ref{Eq-def-e}) and
$\gcd(r, 3^m-1)=2$. Then $\mathcal{C}_{(1,e,s)}$ has no codeword of Hamming weight 3 if $f(x)=x^{r/2}$ is PN over $\fthreem$.
\end{Lemma}

\begin{IEEEproof}
We now prove that (\ref{Eq-(x+1)-w=1+}) has no solution for $\omega=3$. Notice that $\frac{c_ix_i^e}{c_1x_1^e}=\frac{c_i}{c_1}(\frac{c_ix_i}{c_1x_1})^e$ since $e$ is even and $c_i/c_1\in\{1,-1\}$ for $i=2,3$. On the other hand, by $\eta(-1)=-1$, one has $\frac{c_i}{c_1}\eta(\frac{x_i}{x_1})=\eta(\frac{c_ix_i}{c_1x_1})$. Thus, let $x=c_2x_2/c_1x_1$ and $y=c_3x_3/c_1x_1$, then  (\ref{Eq-(x+1)-w=1+}) can be written as
 \begin{equation}\label{Eq-(x+1)-ci-xy}
 \left\{\begin{array}{lll}
   1+x+y=0
  \\ 1+\frac{c_2}{c_1}x^e+\frac{c_3}{c_1}y^e=0
  \\ 1+\eta(x)+\eta(y)=0.
  \end{array} \right.
 \end{equation}
 Assume that $(x,y)$ is a solution of (\ref{Eq-(x+1)-ci-xy}), then by $1+\eta(x)+\eta(y)=0$, one has $\eta(x)=\eta(y)=1$.
 It then follows  from $\gcd(r, 3^m-1)=2$ that there exist $u, v\in \fthreem$ such that $x=u^r$ and $y=v^r$. Then, according to (\ref{Eq-def-e}), one gets $x^e=u^{er}=u^{2\cdot 3^{\tau}}$ and  $y^e=v^{er}=v^{2\cdot 3^{\tau}}$. Therefore,  (\ref{Eq-(x+1)-ci-xy}) is equivalent to
  \begin{equation}\label{Eq-(x+1)-ci-uv}
 \left\{\begin{array}{lll}
   1+u^r+v^r=0
  \\ 1+\frac{c_2}{c_1}u^2+\frac{c_3}{c_1}v^2=0.
  \end{array} \right.
 \end{equation}
 Notice that $r/2$ is even since $f(x)=x^{r/2}$ is PN over $\fthreem$. This leads to $u^r=(\frac{c_2}{c_1}u^2)^{r/2}$ and $v^r=(\frac{c_3}{c_1}v^2)^{r/2}$. Let $\mu=\frac{c_2}{c_1}u^2$ and $\nu=\frac{c_3}{c_1}v^2$. Then (\ref{Eq-(x+1)-ci-uv}) can be written as
 \[f(\mu)-f(1)=f(1)-f(\nu),\;\mu-1=1-\nu\]
 which hold if and only if $(\mu,1)=(1,\nu)$ according to (\ref{Eq-PN-property}) since $f(x)=x^{r/2}$ is PN over $\fthreem$. However, $\mu=\nu=1$ implies  $x=y=1$. This leads to $c_1x_1=c_2x_2=c_3x_3$ which is impossible since $c_1,c_2,c_3\in\{1,-1\}$ and  $x_1,x_2,x_3$ are pairwise distinct. Therefore, (\ref{Eq-(x+1)-ci-xy}) has no solution $(x, y) \in \fthreem^* \times \fthreem^*$.
 This completes the proof.
\end{IEEEproof}

We also need the following lemma in the sequel.

\begin{Lemma}\label{lem-odd-m-w-not=4}
Let $m$ be odd and let $s=\frac{3^m-1}{2}$. Let $e$ and $r$ be even positive integers satisfying (\ref{Eq-def-e}) and
$\gcd(r, 3^m-1)=2$. Then $\mathcal{C}_{(1,e,s)}$ has no codeword of Hamming weight $4$ if $f(x)=x^{r/2}$ is PN over $\fthreem$.
\end{Lemma}

\begin{IEEEproof}
We now show that (\ref{Eq-(x+1)-w=1+}) has no solution for $\omega=4$. Similar as in (\ref{Eq-(x+1)-ci-xy}),  let $x=c_2x_2/c_1x_1$, $y=c_3x_3/c_1x_1$ and $z=c_4x_4/c_1x_1$, then (\ref{Eq-(x+1)-w=1+}) can be written as
 \begin{equation}\label{Eq-(x+1)-ci-xyz}
 \left\{\begin{array}{lll}
   1+x+y+z=0
  \\ 1+\frac{c_2}{c_1}x^e+\frac{c_3}{c_1}y^e+\frac{c_4}{c_1}z^e=0
  \\ 1+\eta(x)+\eta(y)+\eta(z)=0.
  \end{array} \right.
 \end{equation}
Assume that $(x,y,z)$ is a solution of (\ref{Eq-(x+1)-ci-xyz}). Since $1+\eta(x)+\eta(y)+\eta(z)=0$,  we need consider the following three cases: 1) $\eta(x)=1, \eta(y)=\eta(z)=-1$; 2) $\eta(x)=\eta(y)=-1,\eta(z)=1$; and 3)  $\eta(x)=\eta(z)=-1,\eta(y)=1$. In the following, we give only the proof for case 1) since the other two cases can be completely proven in the same manner due to symmetry.

Note that $\eta(-1)=-1$ and $\gcd(r, 3^m-1)=2$. Every square (resp. nonsquare) in $\fthreem$ can be expressed as $u^r$  (resp. $-u^r$) for some $u\in\fthreem$. Thus, for $x,y,z$ with $\eta(x)=1$ and $\eta(y)=\eta(z)=-1$,  there exist $u, v, w\in\fthreem$ such that $x=u^r, y=-v^r$ and $z=-w^r$.  Then, according to (\ref{Eq-def-e}) and the fact that $e$ is even, one gets $x^e=u^{er}=u^{2\cdot 3^{\tau}}$, $y^e=v^{er}=v^{2\cdot 3^{\tau}}$ and $z^e=w^{er}=w^{2\cdot 3^{\tau}}$. Therefore,  (\ref{Eq-(x+1)-ci-xyz}) is equivalent to
    \begin{equation}\label{Eq-(x+1)-ci-uvw-case-1}
     \left\{\begin{array}{lll}
       1+u^r-v^r-w^r=0
     \\ 1+\frac{c_2}{c_1}u^2+\frac{c_3}{c_1}v^2+\frac{c_4}{c_1}w^2=0.
    \end{array} \right.
    \end{equation}
    Let $\mu=\frac{c_2}{c_1}u^2$,  $\nu=\frac{c_3}{c_1}v^2$ and $\lambda=\frac{c_4}{c_1}w^2$. Since $f(x)=x^{r/2}$ is PN over $\fthreem$,  $r/2$ must be even. This implies that $f(\pm\mu)=\mu^{r/2}=(\frac{c_2}{c_1}u^2)^{r/2}=u^r$, $f(\pm\nu)=\nu^{r/2}=v^r$ and $f(\pm\lambda)=\lambda^{r/2}=w^r$. Thus, (\ref{Eq-(x+1)-ci-uvw-case-1}) can be rewritten as
    \begin{equation}\label{Eq-(x+1)-ci-PN-case-1}
     \left\{\begin{array}{ccc}
       f(\mu)-f(-\nu)&=&f(-\lambda)-f(1)
     \\ \mu-(-\nu) &=&(-\lambda)-1.
    \end{array} \right.
    \end{equation}
    Since $f(x)=x^{r/2}$ is PN over $\fthreem$, then by (\ref{Eq-PN-property}) one has that (\ref{Eq-(x+1)-ci-PN-case-1}) holds if and only if $(\mu,-\nu)=(-\lambda, 1)$, i.e., $\mu+\lambda=0$ and $\nu=-1$. This leads to $x+z=0$ and $y=-1$, i.e., $c_2x_2+c_4x_4=0$ and $c_3x_3=-c_1x_1$ which imply that $c_2=c_4$ and $c_1=c_3$ since $x_2\ne x_4$ and $x_1\ne x_3$. Thus, one then has that $x_2=-x_4$, $x_1=-x_3$, and the second equation in (\ref{Eq-(x+1)-w=1+}) can be reduced to $2c_1x_1^e+2c_2x_2^e=0$ which implies that $(2c_1x_1^e)^r=(-2c_2x_2^e)^r$, i.e., $x_1^2=x_2^2$ due to (\ref{Eq-def-e}) and the fact that $r$ is even. This is impossible since $x_1\ne x_2$ and $x_1=-x_2$ implies $x_2=x_3$, a contradiction.  Therefore, (\ref{Eq-(x+1)-ci-xyz})  has no solution over $\fthreem$. This completes the proof.
\end{IEEEproof}

We are now ready to document the main result of this section with the following theorem.

\begin{Theorem}\label{thm-(x+1)-odd-m}
Let $m$ be odd, $e$ be even and $s=\frac{3^m-1}{2}$. Let $r, \tau$ be nonnegative integers such that $\gcd(r, 3^m-1)=2$ and $e\cdot r\equiv 2\cdot 3^{\tau} \pmod{3^m-1}$. Then the ternary cyclic code $\mathcal{C}_{(1,e,s)}$ is optimal and has parameters $[3^m-1,3^m-2m-2, 5]$ if $f(x)=x^{r/2}$ is PN over $\fthreem$.
\end{Theorem}

\begin{IEEEproof}
Since $e$ is even and $e\cdot r\equiv 2\cdot 3^{\tau} \pmod{3^m-1}$, we have $\gcd(e,3^m-1)=2$. It then follows from
Lemma \ref{lem-zero} that $|C_e|=m$. In addition, it is easily verified that $C_1 \cap C_e=\emptyset$. Hence the dimension of the code is equal to $3^m-2m-2$.  The desired conclusion on the minimum distance of this code follows from Lemmas \ref{lem-odd-m-w-not=3}, \ref{lem-odd-m-w-not=4} and Theorem \ref{thm-d<=5}.
 This completes the proof.
\end{IEEEproof}

With Theorem \ref{thm-(x+1)-odd-m}, new classes of optimal ternary cyclic codes $\mathcal{C}_{(1,e,s)}$ with parameters $[3^m-1,3^m-2-2m,5]$ for odd $m$ can be obtained.

\begin{Corollary}\label{cor-e-APN}
 Let $m$ be odd and $s=\frac{3^m-1}{2}$. Then the ternary cyclic code $\mathcal{C}_{(1,e,s)}$ is optimal and has parameters $[3^m-1,3^m-2m-2, 5]$ if
\begin{enumerate}
      \item $e=\frac{3^{m}+1}{4}+\frac{3^m-1}{2}$; or
      \item $e=\frac{3^{m+1}-1}{8}$ for $m\equiv 3 \pmod{4}$; and $e=\frac{3^{m+1}-1}{8}+\frac{3^m-1}{2}$ for $m\equiv 1 \pmod{4}$; or
      \item $e=3^{(m+1)/2}-1$; or
      \item $e=\frac{3^{(m+1)/2}-1}{2}$ for $m\equiv 3 \pmod{4}$; and $e=\frac{3^{(m+1)/2}-1}{2}+\frac{3^m-1}{2}$ for $m\equiv 1 \pmod{4}$; or
      \item $e=(3^{(m+1)/4}-1)(3^{(m+1)/2}+1)$ for $m\equiv 3 \pmod{4}$.
\end{enumerate}
\end{Corollary}

\begin{IEEEproof}
To prove this corollary with Theorem \ref{thm-(x+1)-odd-m}, for each $e$ it is sufficient to find a suitable $r$ such that $e$ and $r$ satisfy the conditions given in Theorem \ref{thm-(x+1)-odd-m}. Thus for the given $e$ in each case, we can define the $r$ respectively by
 \begin{enumerate}
   \item $r=4$;
    \item $r=8$;
   \item $r=3^h+1$, where $h=(m+1)/2$ if $m\equiv 1 \pmod{4}$, and $h=(m-1)/2$ if $m\equiv 3 \pmod{4}$;
   \item $r=2(3^h+1)$, where $h=(m+1)/2$;
   \item $r=3^h+1$, where $h=(m+1)/4$ if $m\equiv 3 \pmod{8}$, and $h=(3m-1)/4$ if $m\equiv 7 \pmod{8}$.
   \end{enumerate}
Then, for each pair $(e,r)$, one can verify that the conditions in Theorem \ref{thm-(x+1)-odd-m} are met according to the known three classes of PN monomials. This completes the proof.
\end{IEEEproof}

Note that all the exponents $e$ given in Corollary \ref{cor-e-APN} are APN exponents \cite{Zha}.

\begin{Example} Let $m=3$, $s=\frac{3^m-1}{2}=13$ and $\alpha$ be a generator of $\fthreem^*$ with $\alpha^3+2\alpha+1=0$.
    \begin{enumerate}
      \item If $e=3^{(m+1)/2}-1=8$, then the generator polynomial of $\mathcal{C}_{(1,8,13)}$ is $x^7+2x^4+x^3+2x+2$ and $\mathcal{C}_{(1,8,13)}$ has parameters $[26,19,5]$.
      \item If $e=\frac{3^{m+1}-1}{8}=10$, then the generator polynomial of $\mathcal{C}_{(1,10,13)}$ is $x^7+2x^6+x^4+2x^2+2$ and $\mathcal{C}_{(1,10,13)}$ has parameters $[26,19,5]$.
    \end{enumerate}
 Both $\mathcal{C}_{(1,8,13)}$ and $\mathcal{C}_{(1,10,13)}$ are optimal according to Corollary \ref{cor-e-APN}.
  \end{Example}

    \begin{Example} Let $m=5$, $s=\frac{3^m-1}{2}=121$ and $\alpha$ be a generator of $\fthreem^*$ with $\alpha^5+2\alpha+1=0$.
    \begin{enumerate}
      \item If $e=\frac{3^{m}+1}{4}+\frac{3^m-1}{2}=182$, then the generator polynomial of $\mathcal{C}_{(1,182,121)}$ is $x^{11}+x^9+x^5+x^4+2x^3+2x^2+2$ and $\mathcal{C}_{(1,182,121)}$ has parameters $[242,231,5]$.
      \item If $e=\frac{3^{(m+1)/2}-1}{2}+\frac{3^m-1}{2}=134$, then the generator polynomial of $\mathcal{C}_{(1,134,121)}$ is $x^{11}+x^8+x^6+2x^5 +2x^4+x^2+x+2$ and $\mathcal{C}_{(1,134,121)}$ has parameters $[242,231,5]$.
    \end{enumerate}
 Both $\mathcal{C}_{(1,182,121)}$ and $\mathcal{C}_{(1,134,121)}$ are optimal according to Corollary \ref{cor-e-APN}.
\end{Example}

 According to Theorem \ref{thm-(x+1)-odd-m}, optimal ternary cyclic code $\mathcal{C}_{(1,e,s)}$  with parameters $[3^m-1,3^m-2-2m,5]$ can also be obtained from $e$ which is not in the list of known APN exponents \cite{Zha}.
The following Corollary is such an example.

\begin{Corollary}\label{cor-e-not-APN}
Let $m\equiv 3 \pmod{4}$, $s=\frac{3^m-1}{2}$ and $e=\frac{3^m+2\cdot 3^t-1}{20}$ with $t\equiv 3 \pmod{4}$.
Then the ternary cyclic code $\mathcal{C}_{(1,e,s)}$ is optimal and  has parameters $[3^m-1,3^m-2m-2, 5]$.
  \end{Corollary}

\begin{IEEEproof}
By a direct calculation, one has that $\frac{3^m-1}{2}\equiv 13 \pmod{20}$ and $3^t\equiv 7 \pmod{20}$ if $m\equiv 3 \pmod{4}$ and $t \equiv 3 \pmod{4}$. Hence $e$ is an even integer. If $r=20$, then $\gcd(r,3^m-1)=2$, $e\cdot r \equiv 2 \cdot 3^t \pmod{3^m-1}$ and $x^{r/2}$ is PN over $\fthreem$ as $m$ is odd. Then the desired result follows from Theorem \ref{thm-(x+1)-odd-m}. This completes the proof.
\end{IEEEproof}

\begin{Example} Let $m=7$, $s=\frac{3^m-1}{2}=1093$ and $\alpha$ be a generator of $\fthreem^*$ with $\alpha^7+2\alpha+1=0$. If $t=3$ and $e=\frac{3^m+2\cdot 3^t-1}{20}=112$ which is not an APN exponent, then the generator polynomial of $\mathcal{C}_{(1,112,1093)}$ is $x^{15}+2x^{14}+2x^{13}+2x^{11}+x^{10}+x^9+2x^8+x^7+x^5 +2x^4+2x^3+x^2+2$ and $\mathcal{C}_{(1,112,1093)}$ has parameters $[2186,2171,5]$ which is optimal according to Corollary \ref{cor-e-not-APN}.
\end{Example}

It should be noticed that Corollary \ref{cor-e-not-APN} in fact indicates a general method for deriving new exponent $e$
such that $\mathcal{C}_{(1,e,s)}$ is optimal and has parameters $[3^m-1,3^m-2m-2, 5]$. In general, with the known PN monomials, one can select an integer $r$ such that $x^{r/2}$ is PN over $\fthreem$ and $\gcd(r, 3^m-1)=2$. Then, for such a fixed $r$, by solving congruence equation (\ref{Eq-def-e}) which has exactly two solutions since $\gcd(r, 3^m-1)=2$, one can get a desired $e$. Hence, through direct calculations, more new optimal ternary cyclic codes $\mathcal{C}_{(1,e,s)}$ with parameters $[3^m-1,3^m-2m-2, 5]$ can be obtained from this approach.

\section{Conclusions}

Optimal ternary cyclic codes $\mathcal{C}_{(1,e)}$ with parameters $[3^m-1,3^m-1-2m,4]$ and $\mathcal{C}_{(1,e,s)}$ with parameters $[3^m-1,3^m-2-2m,5]$ were investigated respectively in this paper. By analyzing irreducible factors of certain polynomials with low degrees over finite fields, an open problem about $\mathcal{C}_{(1,e)}$  for $e=2(3^{m-1}-1)$ proposed by Ding and Helleseth in \cite{Ding-Helleseth} was solved and new optimal ternary codes $\mathcal{C}_{(1,e)}$  with parameters $[3^m-1,3^m-1-2m,4]$ were also obtained with the same techniques.  Moreover, inspired by the work of \cite{Carlet-Ding-Yuan} and \cite{Ding-Helleseth}, a number of classes of optimal ternary cyclic codes $\mathcal{C}_{(1,e,s)}$ with parameters $[3^m-1,3^m-2m-2, 5]$ were also presented. The construction and properties of these optimal codes $\mathcal{C}_{(1,e,s)}$ are based on known PN and APN monomials. Although the length, dimension and the minimum distance of $\mathcal{C}_{(1,e,s)}$ documented in this paper are the same as a class of ternary codes presented in \cite{Carlet-Ding-Yuan}, the codes documented in this paper are different from those in \cite{Carlet-Ding-Yuan} as their generator polynomials are different.


\begin{thebibliography}{99}

 \bibitem{Betti-Sala} E. Betti, M. Sala, ``A new bound for the minimum distance of a cyclic code from its defining set,"
 {\em IEEE Trans. Inform. Theory,} vol. 52, no. 8, pp. 3700--3706, 2006.

 \bibitem{Carlet-Ding-Yuan} C. Carlet, C. Ding, J. Yuan, ``Linear codes from highly nonlinear functions and their secret sharing schemes," {\em IEEE Trans. Inform. Theory,} vol. 51, no. 6, pp. 2089--2102, 2005.

 \bibitem{Coulter-Matthews} R. S. Coulter, R. W. Matthews, ``Planar functions and planes of LenzBarlotti class II,"
 {\em Des. Codes Cryptogr.,} vol. 10, no. 2, pp. 167--184, 1997.

\bibitem{Dembowski-Ostrom} P. Dembowski, T. G. Ostrom, ``Planes of order $n$ with collineation groups of order $n^2$," {\em Math. Z.,} vol. 193, no. 3, pp. 239-258, 1968.

\bibitem{Ding-Helleseth} C. Ding, T. Helleseth, Optimal ternary cyclic codes from monomials, to appear in
{\it IEEE Trans. Inform. Theory}. Availabe at http://arxiv.org/pdf/1305.0061v1.pdf

\bibitem{Ding-Ling} C. Ding, S. Ling, ``A $q$-polynomial approach to cyclic codes,"
  {\em Finite Fields and Their Applications,} vol. 20, no. 3, pp. 1-14, 2013.

\bibitem{Ding-Yang} C. Ding, J. Yang, ``Hamming weights in irreducible cyclic codes," {\em Discrete Mathematics,}
  vol. 313, no. 4, pp. 434--446, 2013.

\bibitem{Rouayheb-GSS} S. Y. El Rouayheb, C. N. Georghiades, E. Soljanin, A. Sprintson, ``Bounds on codes based on graph theory," In: {\em Proc. IEEE Int. Symp. on Information Theory,} pp. 1876-1879, June 2007.

\bibitem{Feng07} K. Feng, J. Luo, ``Weight distribution of some reducible cyclic codes,''
{\em Finite Fields Appl.,} vol. 14, no. 4, pp. 390--409,  2008.

\bibitem{TFeng} T. Feng, ``On cyclic codes of length $2^{2^r}-1$ with two zeros whose
dual codes have three weights,'' {\em Des. Codes Cryptogr.}, vol. 62, no. 3, pp. 253--258, 2012.

\bibitem{Helleseth-Rong-S} T. Helleseth, C. Rong, D. Sandberg, ``New families of almost perfect nonlinear power mappings," {\em IEEE Trans. Inf. Theory,} vol. 45, no. 2, pp. 475--485, 1999.

\bibitem{Jia-Ling-Xing} Y. Jia, S. Ling, C. Xing, ``On self-dual cyclic codes over finite fields,"
  {\em IEEE Trans. Inform. Theory,} vol. 57, no. 4, pp. 2243-2251, 2011.

\bibitem{Lidl-N} R. Lidl, H. Niederreiter, {\em Finite fields,}   Encyclopedia of Mathematics and Its Applications,
vol. 20. Reading, Mass.: Addison-Wesley, 1983.

\bibitem{LuoFeng2} J. Luo, K. Feng, ``On the weight distributions of two classes
of cyclic codes,'' {\em IEEE Trans. Inform. Theory,} vol. 54, no. 12, pp. 5332--5344, 2008.

\bibitem{YCD06} J. Yuan, C. Carlet, C. Ding, ``The weight distribution of a
      class of linear codes from perfect nonlinear functions,''
      {\em IEEE Trans. Inform. Theory}, vol. 52, no. 2, pp. 712--717, Feb. 2006.

 \bibitem{Zeng-HJYC} X. Zeng, L. Hu, W. Jiang, Q. Yue, X. Cao, ``The weight distribution of a class of $p$-ary cyclic codes," {\em Finite Fields and Their Applications,} vol 16, no. 1, pp. 56--73, 2010.

 \bibitem{Zeng-SH} X. Zeng, J. Shan, L. Hu, ``A triple-error-correcting cyclic code from the Gold and Kasami-Welch APN power functions," {\em Finite Fields and Their Applications,} vol. 18, no. 1, pp, 70--92, 2012.

\bibitem{Zha} Z. Zha, X. Wang, ``Almost perfect nonlinear power functions in odd characteristic",
{\em IEEE Trans. Inform. Theory,} vol. 57, no. 7, pp. 4826--4832, 2011.

\end{thebibliography}
\end{document}